%% file: main_arxiv.tex
\documentclass[aps,showpacs,twocolumn,superscriptaddress]{revtex4-2}

\input{macros}  

\usepackage{dcolumn}
\usepackage{booktabs}

\usepackage{subcaption}
\DeclareCaptionJustification{blocksatz}{\flushing}
\newcommand{\tabfont}{\footnotesize}

\usepackage{natbib}
\usepackage[colorlinks=true]{hyperref}
\usepackage[normalem]{ulem}
\hypersetup{
           breaklinks=true,
           colorlinks=true,
           pdfusetitle=true, 
           citecolor=blue,
           urlcolor=blue
           }

\begin{document}
\title{Circuit Design based on Feature Similarity for Quantum Generative Modeling}

\author{Mathis Makarski}
\email{makarski@keio.jp}
\affiliation{Graduate School of Science and Technology, Keio University, Hiyoshi 3-14-1, Kohoku, Yokohama, Kanagawa 223-8522, Japan}
\affiliation{RWTH Aachen University, Aachen, Germany}

\author{Jumpei Kato}
\email{jumpei\_kato@keio.jp}
\affiliation{Mitsubishi UFJ Financial Group, Inc.~and MUFG Bank, Ltd., 4-10-2 Nakano, Nakano-ku, Tokyo 164-0001, Japan}
\affiliation{Quantum Computing Center, Keio University, Hiyoshi 3-14-1, Kohoku, Yokohama, Kanagawa 223-8522, Japan}
\affiliation{Graduate School of Science and Technology, Keio University, Hiyoshi 3-14-1, Kohoku, Yokohama, Kanagawa 223-8522, Japan}

\author{Yuki Sato}
\affiliation{Toyota Central R\&D Labs., Inc., 1-4-14 Koraku, Bunkyo-ku, Tokyo 112-0004, Japan}
\affiliation{Quantum Computing Center, Keio University, Hiyoshi 3-14-1, Kohoku, Yokohama, Kanagawa 223-8522, Japan}

\author{Naoki Yamamoto}
\affiliation{Quantum Computing Center, Keio University, Hiyoshi 3-14-1, Kohoku, Yokohama, Kanagawa 223-8522, Japan}
\affiliation{Department of Applied Physics and Physico-Informatics, Keio University, Hiyoshi 3-14-1, Kohoku, Yokohama, Kanagawa 223-8522, Japan}

\begin{abstract}
\input{content/0_abstract}
\end{abstract}

\maketitle

\input{content/1_introduction}
\input{content/2_methods}
\input{content/3_results}
\input{content/4_discussion}

\begin{acknowledgments}
\input{content/acks}
\end{acknowledgments}

\bibliography{ref_v2}

\appendix
\input{content/appendix}

\end{document}

%% file: macros.tex
\usepackage{braket}
\usepackage{physics}
\usepackage{bm}
\usepackage{graphicx}
\usepackage{amsfonts,amsmath,amssymb,amsthm}

%% file: content/0_abstract.tex
Quantum generative models are among the most promising candidates for a practical application of near-term quantum devices. However, current approaches mostly rely on general-purpose circuits, such as the hardware efficient ansatz paired with a random initialization strategy, which are known to suffer from trainability issues such as barren plateaus. To address these issues, a tensor network pretraining framework that initializes a quantum circuit ansatz with a classically computed high-quality solution for a linear entanglement structure has been proposed in literature. In order to improve the classical solution, the quantum circuit needs to be extended, while it is still an open question how the extension affects trainability. In this work, we propose the metric-based extension heuristic to design an extended circuit based on a similarity metric measured between the dataset features. We evaluate this method on the bars and stripes dataset and carry out simulations on financial data. Our results provide an experimental basis for problem-informed circuit design and show that the metric-based extension heuristic offers the means to introduce inductive bias while designing a circuit under limited resources.

%% file: content/1_introduction.tex
\section{Introduction}
The rise of quantum computing is opening up new possibilities for solving classically intractable problems~\cite{harrow_quantum_2017, huang_quantum_2022, daley_practical_2022, farhi_qaoa_2019}. Despite significant advancements in the theoretical exploration of quantum advantages in terms of speed-up~\cite{harrow_linear_2009, lloyd_quantum_2014, huang_power_2021, huang_quantum_2022}, memory efficiency~\cite{anschuetz_interpretable_2023}, or model expressiveness~\cite{coyle_born_2020, sweke_quantum_2021, gao_quantumcorr_2022, gili_generalize_2023}, the widespread adoption of quantum systems will ultimately depend on their ability to solve problems of high practical interest on hardware with limited resources~\cite{aaronson_read_2015, preskill_nisq_2018, national_academies_of_sciences_quantum_2019, bharti_nisq_2022}. In search of an application of such limited near- and mid-term quantum devices, quantum machine learning~(QML) has emerged as a promising domain~\cite{schuld_quest_2014, schuld_introduction_2015, biamonte_quantum_2017}.

The inherently probabilistic nature of quantum systems potentially allows for efficient sampling from probability distributions of high complexity~\cite{terhal_adaptive_2004, aaronson_linear_optics_2010, farhi_qaoa_2019, arute_quantum_2019, madsen_quantum_2022}, making them well suited for the task of generative modeling within the field of QML~\cite{perdomo_ortiz_opportunities_2018}. Generative models aim to learn the underlying distribution of a dataset in order to generate realistic samples~\cite{bond_taylor_deep_2022}. Recent advances in quantum generative modeling~\cite{tian_recent_2022} have led to the adoption of several architectures, including quantum circuit Born machines~(QCBMs)~\cite{benedetti_qcbm_benchmark_2019, liu_differentiable_2018, coyle_born_2020}, quantum generative adversarial networks~(QGANs)~\cite{lloyd_qgan_2018, dallaire_demers_qgan_2018}, and quantum Boltzmann machines~(QBMs)~\cite{amin_qbm_2018, zoufal_variational_2021}.

While these models provide high expressiveness, rooted in their general-purpose circuit architecture, they suffer from prominent trainability issues, such as cost function concentration and barren plateaus~\cite{mcclean_barren_2018, arrasmith_barren_2022, larocca_diagnosing_2022, cerezo_higher_2021, arrasmith_effect_2021, holmes_scramblers_2021, zhao_barren_2021, thanasilp_exponential_2024, ragone_lie_2024, fontana_characterizing_2024, anschuetz_quantum_2022}. When this occurs, the loss function values exponentially concentrate around a fixed value and the loss gradients become exponentially flat with growing problem size, therefore requiring exponential resources to navigate through the loss landscape. This phenomenon originates from several sources~\cite{cerezo_challenges_2022}, including insufficient inductive bias~\cite{holmes_connecting_2022}, global observables~\cite{cerezo_cost_2021, uvarov_barren_2021}, and too much entanglement~\cite{sharma_trainability_2022, marrero_entanglement_2021, patti_entanglement_2021}. Moreover, the quantum no-free-lunch theorem implies that problem-agnostic approaches, such as the QCBM with a hardware-efficient ansatz and random initialization, tend to exhibit poor average performance~\cite{ho_no_free_lunch_2002, poland_no_free_lunch_2020, sharma_no_free_lunch_2022}. To mitigate these trainability issues, various techniques have been proposed, such as sophisticated parameter initialization strategies~\cite{verdon_learning_2019, grant_initialization_2019, rudolph_synergistic_2023} and problem-informed circuit design~\cite{larocca_group_2022, meyer_exploiting_2023, zheng_speeding_2023, bowles_contextuality_2023}.

In this work, we focus on providing experimental results for techniques that are shown to improve trainability on near-term quantum devices. We combine both strategies of initializing the parameters of a general-purpose ansatz by tensor network pretraining, and further extending it to a problem-informed circuit based on classical feature similarity. More specifically, we employ the synergistic pretraining framework~\cite{rudolph_synergistic_2023} that utilizes matrix product states~(MPS)~\cite{han_unsupervised_2018} to obtain a high-quality solution for a given problem, which can then be transferred to a quantum circuit ansatz with linear connectivity~\cite{rudolph_decomposition_2023}. This approach offers a hybrid solution for parameter initialization that is classically scalable to larger qubit numbers, due to its restriction to a linear entanglement structure. However, to achieve a potential performance increase over this classically pretrained solution, the quantum circuit needs to be extended. The goal is to let the extended circuit represent non-linear correlations that are difficult to capture effectively by an MPS with modest bond dimensions. It has been empirically shown that, in contrast to a random initialization strategy, the loss gradients of pretrained circuits -- whether left unextended or extended to the trivial all-to-all connectivity -- exhibit a beneficial scaling behavior for larger qubit circuits~\cite{rudolph_synergistic_2023}. Although the problem-agnostic all-to-all circuit extension exhibits this beneficial scaling in~\cite{rudolph_synergistic_2023}, its gate count and compute time still scale unfavorably, quickly rendering it infeasible to train on hardware with limited resources. Given the disadvantages of problem-agnostic approaches, we aim for a practical problem-informed approach to extend the pretrained circuit. 
 
In recent QML literature, probabilistic graphical models~(PGMs)~\cite{koller_probabilistic_2009}, especially Bayesian networks~\cite{low_quantum_inference_2014, borujeni_bayesian_2021, gao_quantumcorr_2022} and Markov networks~\cite{bako_problem_informed_2024} are being exploited to construct problem-informed QML models, with the aim of creating new classes of quantum algorithms that potentially demonstrate advantages over classical algorithms. In contrast to the works that focus on creating new classes of algorithms, our approach is more practically motivated, similar to Liu and Wang~\cite{liu_differentiable_2018}, who made the first explicit connection between QCBMs and PGMs. In order to construct a circuit that exploits the close relation between controlled unitary gates and classical PGMs~\cite{low_quantum_inference_2014}, they employed the Chow-Liu dependency tree~\cite{chow_liu_tree_1968}. By maximizing mutual information, the Chow-Liu tree approximates a high-dimensional joint distribution through a product of second-order distributions. However, being a second-order product approximation, this method not only fails to capture higher-order correlations but also cannot represent cyclical dependencies. 

The main contribution of this work is the introduction of the metric-based extension heuristic to design a problem-informed circuit connectivity that can be used to extend the classically pretrained ansatz in a greedy fashion. Being a practically motivated heuristic, this method might not offer an optimal solution. However, it does not suffer from the drawbacks of the Chow-Liu tree and provides the means for finding a suitable tradeoff between the number of parameters and the amount of inductive bias. Therefore, our heuristic occupies a practical middle ground between the Chow-Liu tree~\cite{chow_liu_tree_1968} and the all-to-all connected circuit proposed in~\cite{rudolph_synergistic_2023}. As we aim for a flexible method that is also applicable to real-world datasets, we apply the metric-based extension heuristic not only to a toy problem, the bars and stripes~(BAS) dataset, but also to a relevant financial dataset composed of Japanese government bond~(JGB) interest rates. We find that our heuristic generally outperforms randomly extended circuits with the same number of trainable parameters, and compares favorably to the all-to-all connected circuit at a fraction of its parameter cost. Although the employed algorithm may still face trainability and generalization issues, our findings indicate that the metric-based extension heuristic introduces useful inductive bias, while keeping the resulting quantum circuit shallow.

The remainder of this paper is structured as follows: In Sec.~\ref{sec:methods}, we review relevant methods and introduce our extension heuristic. Sec.~\ref{sec:num_sim} presents the results of the BAS and JGB experiments. Finally, we conclude this work with a discussion of the results and future directions in Sec.~\ref{sec:discussion}.  

%% file: content/2_methods.tex
\section{Methods}\label{sec:methods}

\subsection{Quantum Circuit Born Machine}
QCBMs are a subclass of parameterized quantum circuits~(PQCs) that is used for generative modeling. They contain a set of parameters, which are adjusted during a training process according to a loss function $\mathcal{L}(\theta)$. The QCBM aims to encode the target joint distribution $p(y)$ of a discrete dataset $y \in \{1,0\}^n$ into an $n$-qubit parameterized quantum state $\ket{\psi_{\theta}}$, which is, without loss of generality, evolved from the all zero input state $\ket{0}^{\otimes n}$, i.e., $\ket{\psi_{\theta}} = U(\theta) \ket{0}^{\otimes n}$ using a PQC $U(\theta)$. Measuring the output state yields $x \in \{1,0\}^n$ with $x$ sampled from the variational model distribution $q_\theta  = |\Braket{x | \psi_\theta}|^2$, which is called the Born rule. Unlike in classical models, one does not have access to the explicit model distribution. Therefore, the loss function has to be estimated using quantum circuit measurements. Since the widely used Kullback-Leibler~(KL) divergence, as an explicit loss function, requires a high number of samples, and is shown not to be trainable for large-scale QCBMs in Ref.~\cite{rudolph_trainability_2024}, we use the implicit loss function, maximum mean discrepancy (MMD)~\cite{anderson_mmd_1994}
\begin{equation}\label{eqn:mmd-with-kernel}
    \mathcal{L}_{\rm MMD} = \underset{\substack{x \sim q_\theta \\ y \sim q_\theta}}{\mathbb{E}}(\kappa(x,y)) + \underset{\substack{x \sim p \\ y \sim p}}{\mathbb{E}}(\kappa(x,y)) - \underset{\substack{x \sim q_\theta \\ y \sim p}}{2\mathbb{E}}(\kappa(x,y)),
\end{equation}
where $\kappa(x,y)$ is a freely chosen kernel function~\cite{hofmann_kernel_2008}. As a standard choice in recent literature~\cite{liu_differentiable_2018, coyle_born_2020, rudolph_trainability_2024}, we employ the Gaussian mixture kernel defined as
\begin{equation}\label{eqn:kernel}
    \kappa_G(x,y) = \frac{1}{c} \sum^c_{i=1} \exp \Bigg(-\frac{||x-y||^2_2}{2\sigma_i} \Bigg).
\end{equation}
The parameters $\sigma_i$ are the bandwidths to adjust the width of the Gaussian kernel and $\| \cdot \|^2_2$ denotes the squared $l_2$-norm. Regarding the choice of optimizer, both gradient-free and gradient-based procedures have been proposed, although gradient-free schemes failed when scaling to a larger number of parameters~\cite{benedetti_qcbm_benchmark_2019}. Therefore, we use the stochastic gradient descent~(SGD) algorithm~\cite{sweke_stochastic_2020} and ADAM as meta-optimizer~\cite{diederik_adam_2014}. For the MMD loss, it is possible to obtain the gradient with respect to the $k$-th model parameter $\theta_k$ as
\begin{align}
    \label{eqn:mmd-grad}
    \frac{\partial \mathcal{L}_{\rm MMD}}{\partial \theta_k}
    =& \underset{\substack{a \sim q_{\theta_k}^{-} \\ x \sim q_\theta}}{\mathbb{E}}(\kappa(a,x))
    - \underset{\substack{b \sim q_{\theta_k}^{+} \\ x \sim q_\theta}}{\mathbb{E}}(\kappa(b,x)) \notag \\
    &- \underset{\substack{a \sim q_{\theta_k}^{-} \\ y \sim p}}{\mathbb{E}}(\kappa(a,y))
    + \underset{\substack{b \sim q_{\theta_k}^{+} \\ y \sim p}}{\mathbb{E}}(\kappa(b,y))
\end{align}
by shifting $\theta_k$ and performing a projective measurement to obtain samples. The distributions $q_{\theta_k}^{\pm}$ are generated by sampling from the parameter-shifted circuit with parameters $\theta^{\pm}_k = \theta_k \pm \frac{\pi}{2}$~\cite{schuld_evaluating_2019}.

This optimization procedure is hybrid quantum-classical: the quantum device is used only to obtain measurement samples from the parameter-shifted circuits, while the resulting gradient (Eq.~\ref{eqn:mmd-grad}), the SGD update, and ADAM's adaptation of the SGD learning rate are all computed entirely on a classical device. Each iteration measures the circuit at the current parameters, updates them classically, and repeats until the loss converges.

\subsection{Tensor Network Pretraining}

\begin{figure}[ht]
    \centering{\includegraphics[width=1.0\columnwidth]{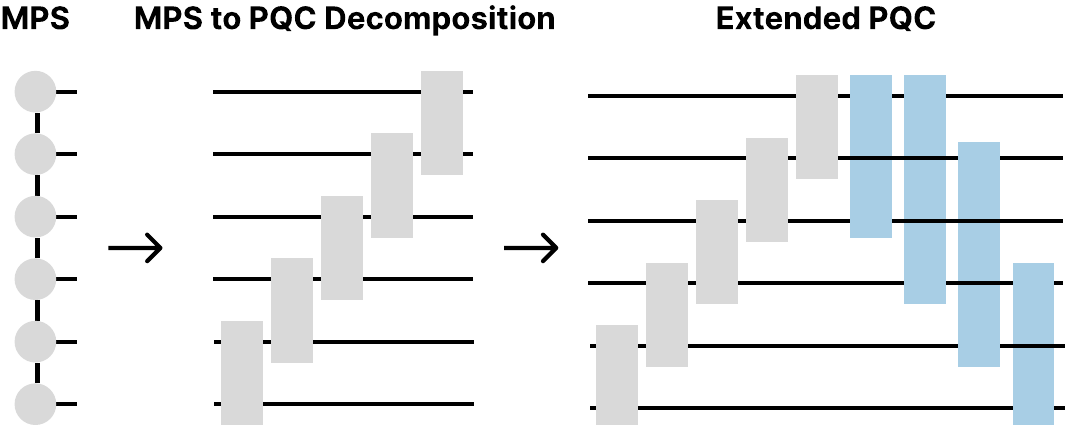}}
    \caption{Tensor network pretraining framework. The circuit from MPS decomposition in gray represents the baseline with linear connectivity. We extend it with additional gates in light blue.}
    \label{fig:pretraining_schematic}
\end{figure}

In the field of quantum mechanics, the development of efficient classical representations of quantum wave functions has been a subject of extensive research, whereby tensor networks (TN) have demonstrated notable effectiveness. Matrix product states (MPS) are a particular type of TN, whose tensors are arranged in a one-dimensional geometry, facilitating the representation of one-dimensionally entangled quantum states. The MPS provides means for efficient computation of quantum circuits with limited entanglement~\cite{vidal_efficient_2003} via the density matrix renormalization group (DMRG) algorithm~\cite{schollwock_density_matrix_2005}. Although the original application of DMRG is ground state computation, it has also been successfully applied to unsupervised learning tasks~\cite{han_unsupervised_2018}, which enables the use of MPS as a pretraining method for quantum generative modeling. In this work, we implemented the synergistic pretraining framework~\cite{rudolph_synergistic_2023}, as schematically shown in Fig.~\ref{fig:pretraining_schematic}, to mitigate the problem of barren plateaus in random initialization.

First of all, the MPS is classically trained to find a high-quality solution by a DMRG-like algorithm that minimizes negative log-likelihood~\cite{han_unsupervised_2018, han_mps_github_2018}. Theoretically, any probability distribution of an $n$-bit system can be represented through an MPS using exponentially large bond dimensions. By removing the components with small singular values in the MPS, we can obtain an approximate MPS, which is known to be locally optimal across neighboring matrices in the MPS~\cite{paeckel2019time}. This truncation corresponds to dropping minor components of the entanglement spectrum. In the present study, we did not explicitly set the upper bound of the bond dimensions, but instead we used a truncation cutoff of $5\times10^{-5}$ to truncate the bond dimensions. We set the training iterations to 10 and the learning rate to $0.05$. In both of our experiments in Sec.~\ref{sec:bas_experiment} and Sec.~\ref{sec:jgb_experiment}, the training concluded after 10 iterations with a realized maximum bond dimension of 5 for the bars and stripes dataset and 25 for the financial dataset. After classical pretraining, the obtained MPS solution is then transferred to a quantum circuit ansatz by analytical decomposition~\cite{rudolph_decomposition_2023, ran_analytic_decomp_2020}. In order to introduce inductive bias to the model, the next step consists of extending the decomposed quantum circuit by additional gates that change its connectivity, therefore changing its entanglement capabilities. Subsequently, the extended circuit should be retrained on quantum hardware. In this work, we explore the effect of different circuit extensions on this subsequent training process. However, we use numerical simulations instead of real quantum hardware and leave this to future work.

\subsection{Metric-based Extension Heuristic}\label{sec:extension-heuristic}

Since the MPS achieves a high-quality solution only for a linear entanglement structure, we do not expect retraining the baseline circuit with limited additional entanglement to significantly improve upon it, even on quantum hardware. To learn long-range correlations, it is necessary to introduce more entanglement capacity, which raises the question of how to extend the baseline circuit in a way that uses a minimal number of gates. For small datasets, connecting the qubits in an all-to-all fashion provided good results~\cite{rudolph_synergistic_2023}. However, the number of parameters grows large with increasing qubits, which leads to deep circuits that are infeasible to train. Instead of all-to-all connections, we introduce a heuristic we term the \textit{metric-based extension heuristic} to reduce the gate count and introduce inductive bias by utilizing characteristics of the dataset.

\subsubsection{Extension Procedure}
The metric-based extension heuristic is carried out in the following steps:

\begin{enumerate}
    \item Construct an undirected graph $G = (V, E)$, where vertices $V$ represent binary features and edges $E$ represent their pair-wise similarity. From the perspective of the circuit, vertices represent the qubit registers, and edges represent a connection by a two-qubit gate. The initial graph resembles the linear connectivity of the decomposed MPS circuit.
    \item Compute the similarity metric as per Sec.~\ref{sec:metric} between each pair of binary feature vectors.
    \item Apply a threshold to the similarity metric to decide whether to add an edge to the graph or not. If the binary features share a high similarity, their representative qubits should therefore be connected.
    \item To extend the linear MPS circuit, construct a new quantum circuit by adding extension gates, as in Sec.~\ref{sec:extension_gates}, between qubits that are connected in the graph.
    \item Initialize the extension gates as near-identity matrices by choosing random parameters from a normal distribution with mean $\mu=0.0$ and small standard deviation, which we set to $\sigma=0.01$.
    \item Retrain the whole circuit including the parameters of the previously derived MPS solution that formed the linear baseline circuit.
\end{enumerate}
Following this procedure, we attempt to design a problem-informed circuit in a greedy way, since setting a higher threshold increases the number of extension gates in a way that adds inductive bias up to a certain point. However, as the threshold is chosen too high, the graph comes close to an all-to-all connectivity and the inductive bias will be lost. To identify a suitable number of extension gates, we apply the so-called ``knee'' (or ``elbow'') heuristic (see Fig.~\ref{fig:bas_threshold_curve} and Fig.~\ref{fig:jgb_threshold}). Nevertheless, a practitioner could also experiment with different thresholds and similarity metrics to find a configuration for a given dataset that results in a sufficient number of extension gates, but also keeps the circuit shallow enough to be trainable.

\subsubsection{Similarity Metric}\label{sec:metric}
The similarity metric is computed between each pair of the $f$ feature vectors from the training dataset $X = [X_0, X_1, ..., X_{f-1}]$, each including the same number of samples. While in principle any metric that measures a similarity between two feature vectors would serve, we propose to use the variation of information~\cite{arabie_varinfo_1973}. In scenarios where nonlinearity is prevalent, this metric turns out to be a more appropriate distance metric than the correlation measure. It allows us to address questions about the unique information provided by a random variable, all without imposing specific functional assumptions. The variation of information $VI$ and its standardized value $\widetilde{VI}$ are defined as
\begin{align}\label{eqn:vi}
    VI(X,Y) = H(X|Y) + H(Y|X), \notag \\
    \widetilde{VI}(X,Y) = \frac{VI(X,Y)}{H(X,Y)},
\end{align}
where $H(X|Y)$, $H(Y|X)$ denote conditional entropies, $H(X,Y)$ denotes the joint entropy, and $X, Y$ are discrete random variables. $VI$ and $\widetilde{VI}$ are metrics because they satisfy nonnegativity, symmetry $VI(X,Y) = VI(Y,X)$, and the triangle inequality. Further, $\widetilde{VI}$ is bounded as $\widetilde{VI} \in [0,1]$. The variation of information measure can be interpreted as the uncertainty we expect in one variable when we know the value of the other. Therefore, a lower value indicates that more information is shared between both variables.

\subsubsection{Extension Gates}\label{sec:extension_gates}
For extending the MPS circuit, we use SU(4) gates between qubit pairs with 15 parameters per gate as in Ref.~\cite{rudolph_synergistic_2023}, which represent a fully parameterized two-qubit interaction. The SU(4) gate can be decomposed into four single-qubit U(2) rotations, and the RXX, RYY, RZZ entanglement gates by KAK-decomposition~\cite{tucci_kak_2005};
\begin{align}
\mathrm{SU}(4)_{i,j}(\bm{\theta}) = & \mathrm{U}(2)_i(\theta_{0:2}) \times \mathrm{U}(2)_j(\theta_{3:5}) \notag \\
& \times \mathrm{RXX}_{i,j}(\theta_6) \times \mathrm{RYY}_{i,j}(\theta_7) \times \mathrm{RZZ}_{i,j}(\theta_8) \notag \\
& \times \mathrm{U}(2)_i(\theta_{9:11}) \times \mathrm{U}(2)_j(\theta_{12:14})
\end{align}
where $\bm{\theta}$ is a parameter vector of length 15, and each U(2) gate consists of three parameters, with one parameter assigned to each entangling gate.

%% file: content/3_results.tex
\section{Numerical Simulations}\label{sec:num_sim}

In the following sections~\ref{sec:bas_experiment} and \ref{sec:jgb_experiment}, we compare our proposal on two practical datasets, bars-and-stripes~(BAS) and interest rates of Japanese government bonds~(JGB).

\subsection{Bars and Stripes}\label{sec:bas_experiment}
\begin{figure}[ht]
\centering{\includegraphics[width=1.0\columnwidth]{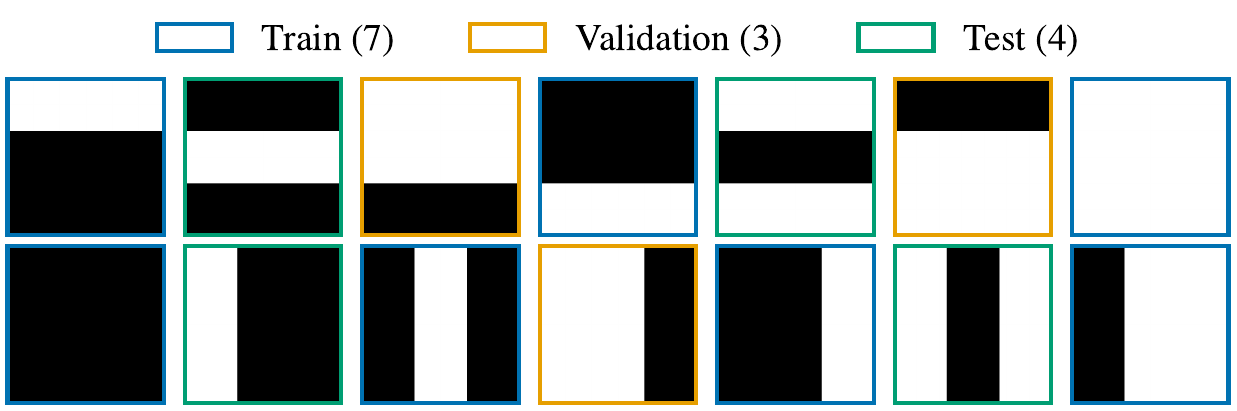}}
\caption{The 14 valid patterns of the $3\times3$ bars-and-stripes (BAS) dataset, each shown as a binary pixel grid (white = 1, black = 0). Border color indicates train/validation/test split membership, with per-split counts given in the legend.}
\label{fig:bas_dataset}
\end{figure}

The BAS dataset is composed of binary images, each containing either vertical bars or horizontal stripes, but never both. The valid BAS images all appear with equal probability and thus resemble a uniform target distribution. Here we use a dimension of $3 \times 3$, resulting in $2^9$ different configurations. The 14 valid images are shown in Fig.~\ref{fig:bas_dataset}. The BAS dataset requires the generative model to capture strong non-trivial correlations and can be scaled to different grid sizes, making it suitable for evaluating the capabilities of quantum generative models with a low number of qubits. Therefore, it is widely used for benchmarking generative modeling tasks in quantum computing literature~\cite{benedetti_qcbm_benchmark_2019, hamilton_benchmark_2019, du_expressive_2020}.

\subsubsection{Circuit Design}

\begin{figure}[ht]
\centering{\includegraphics[width=1.0\columnwidth]{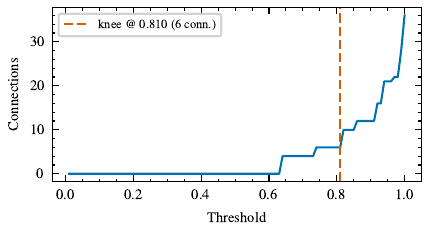}}
\caption{Number of added connections vs.\ threshold for the metric-based extension on the $3\times3$ BAS dataset, using the variation-of-information metric between qubits. The dashed line marks the knee-selected threshold, admitting $6$ connections, two of which are already included in the linear circuit.}
\label{fig:bas_threshold_curve}
\end{figure}

By applying the metric-based extension heuristic, we extend the linear baseline circuit with additional gates. The extension is set by the knee rule on the variation-of-information metric, as shown in Fig.~\ref{fig:bas_threshold_curve}. The resulting metric-based circuit, together with the other connectivities we compare against, is shown in Fig.~\ref{fig:bas_extensions_mmd}. The order of initial linear connections that result from decomposing the MPS can be chosen freely. For the BAS dataset, we therefore swapped the features \{3,5\} to prevent diagonal connections on the two-dimensional grid representing the BAS image. As mentioned earlier, the metric-based extension is conceptually similar to the Chow-Liu method, and therefore shows a nearly identical connectivity. Metric-based amounts to 4 additional gates, whereas Chow-Liu needs 5. With 15 trainable parameters per gate, this results in 60 and 75 parameters, respectively. We further compare the metric-based extension to random connections with the same number of additional parameters. The problem-agnostic all-to-all extension connects every qubit pairwise, requiring 28 extension gates, or 420 additional parameters.

\subsubsection{Training}\label{sec:bas_training}

The experiment was conducted using the IBM Qiskit Aer statevector simulator and consisted of 16 independent training runs per connectivity, all sharing the same initial conditions. The configuration consists of nine qubits and $N_\mathrm{shots} = 2000$ shots per circuit evaluation. For a fair comparison, we allocate a fixed budget of $1 \times 10^{10}$ measurements to each connectivity. With $P$ as the number of trainable circuit parameters, each training iteration executes $(2P+1) \times N_\mathrm{shots}$ measurements via the parameter shift rule. Therefore, the connectivities with a low number of parameters are able to perform more training iterations than those with a high parameter count. To compute the MMD loss function (Eq.~\ref{eqn:mmd-with-kernel}) and its gradient (Eq.~\ref{eqn:mmd-grad}), we used the full training set in each iteration with multiple MMD kernel bandwidths of $\sigma = \{0.25, 1.0, 4.0, 16.0\}$. For a comprehensive collection of hyperparameters, we refer to Appendix~\ref{app:hyperparameters}. 

\begin{figure*}[ht]
\centering
\includegraphics[width=1.0\textwidth]{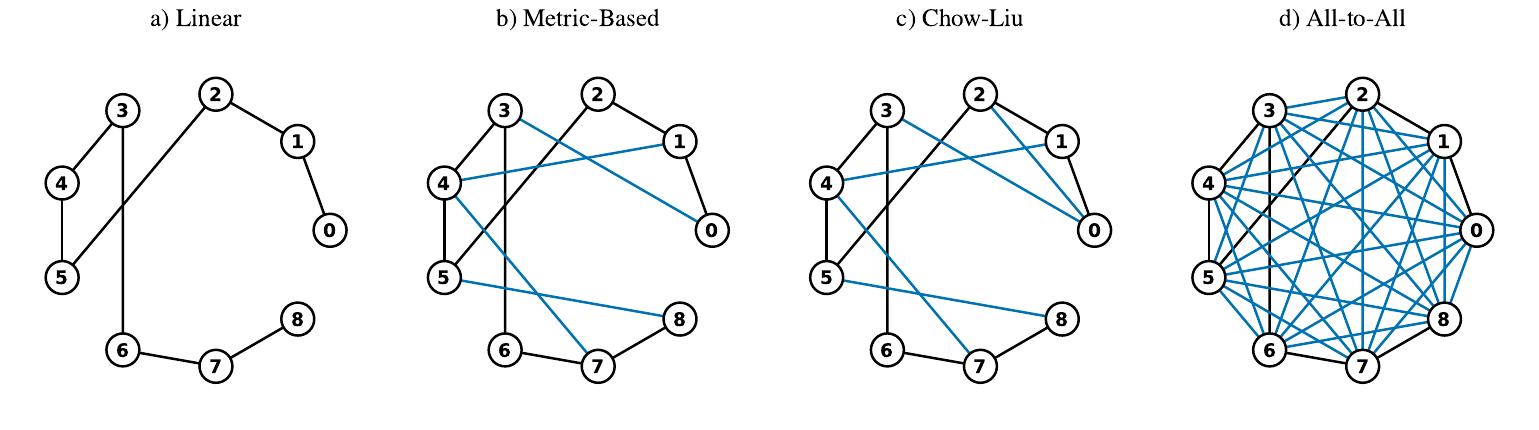}
\includegraphics[width=1.0\textwidth]{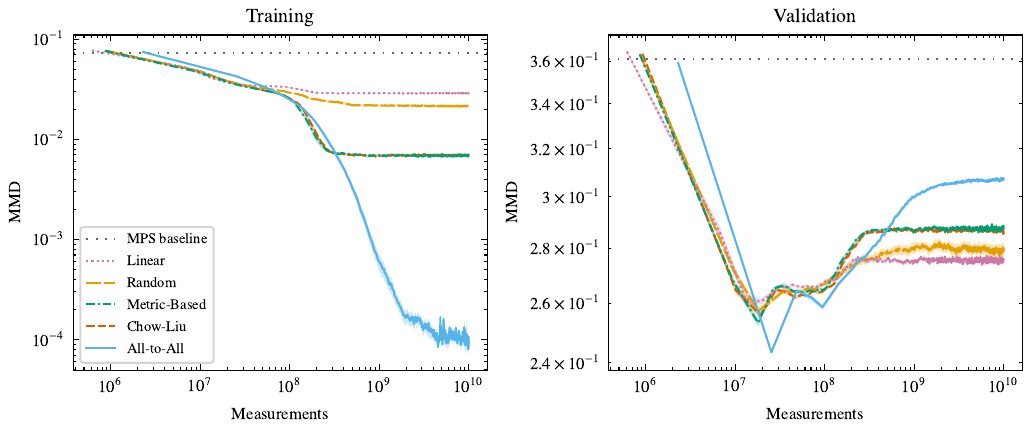}
\caption{Top: connectivity of the BAS circuit extensions. Black edges are the linear connectivity derived from the pretrained MPS model; blue edges are the added connections for (a) no extension, (b) metric-based, (c) Chow-Liu, and (d) all-to-all. Qubits are assigned to the BAS image in row-major order; the 0th qubit corresponds to the top-left cell and the 8th qubit to the bottom-right. Bottom: MMD vs.\ cumulative measurements on the BAS training (left) and validation (right) splits, log-log scale, for the same connectivities: linear, Chow-Liu, metric-based, random, and all-to-all. Lines show the bootstrapped mean and shaded bands show its standard deviation across 16 independent runs per connectivity; the grey dotted line is the MMD of the untrained, MPS-derived linear circuit.
\label{fig:bas_extensions_mmd}}
\end{figure*}

In Fig.~\ref{fig:bas_extensions_mmd}, we compare the MMD loss of different circuit designs on the training and validation sets. The MMD values for both sets were recorded after every tenth training iteration to validate the model performance during the training process. On the training set, the MMD loss separates clearly by connectivity. The all-to-all circuit reaches the lowest loss, whereas metric-based, Chow-Liu and random extensions fall in between. The loss curves of metric-based and Chow-Liu share a nearly identical shape, since both extensions consist of the same connections, with the exception of one. A notable observation is the separation between random connections and metric-based. As both methods share the same number of parameters, this finding indicates that problem-informed circuits are more trainable.

On the validation set, every connectivity instead reaches a plateau close to the MPS baseline, whereas all-to-all shows the highest and linear the lowest MMD values, a sign of overfitting. Denser connectivity leads to a better fit to the training distribution, but that effect does not transfer to the held-out validation set. However, the validation set consists of a very limited number of images, so these results should be read with caution. For that reason, we use the models at the final training iteration for the subsequent benchmarking, instead of the model that performed best on the validation set.

\subsubsection{Benchmarking}
All trained models are benchmarked with $10\,000$ measurements on the full BAS dataset, now also including the previously held-out test set. The benchmarking results in Table~\ref{tab:bas_benchmark} indicate an expressivity-generalization tradeoff across connectivities. The two problem-informed, sparse extensions, Chow-Liu and metric-based, add only 4--5 connections yet attain the lowest MMD, KL, and TV of all connectivities considered. All-to-all achieves the best distributional fidelity and BAS precision by concentrating most of its probability mass on a narrow set of valid images, but this comes at the cost of the worst coverage, exploration, and BAS recall, which is consistent with the overfitting behavior seen in the validation MMD curves (Fig.~\ref{fig:bas_extensions_mmd}). Linear shows the opposite pattern: broad exploration and high BAS recall but the weakest distributional fit and precision. Chow-Liu and metric-based occupy the middle ground, matching most of all-to-all's distributional performance at a fraction of its parameter cost, without a collapse in generalization.

\begin{table*}[ht]
\caption{Bootstrapped benchmark metrics (mean $\pm$ standard error, 16 models per connectivity) on the BAS dataset. Bold marks the best value per metric row, respecting each metric's own direction (lowest for MMD/KL/TV, highest otherwise).}
\label{tab:bas_benchmark}
\tabfont
\begin{tabular*}{\textwidth}{@{\extracolsep{\fill}}lccccc@{}}
\toprule
Metric & Linear & Random & Metric-Based & Chow-Liu & All-to-All \\
\midrule
\multicolumn{6}{l}{\textit{Setup}} \\
parameters       & 153 & 213 & 213 & 228 & 573 \\
extension gates  & 0 & 4 & 4 & 5 & 28 \\
measurements & 9{,}999{,}604{,}000 & 9{,}999{,}486{,}000 & 9{,}999{,}486{,}000 & 9{,}999{,}160{,}000 & 9{,}999{,}546{,}000 \\
iterations   & 16{,}286 & 11{,}709 & 11{,}709 & 10{,}940 & 4{,}359 \\
\midrule
\multicolumn{6}{l}{\textit{Distribution distance metrics}} \\
MMD       & 0.0424 $\pm$ 0.0003 & 0.0420 $\pm$ 0.0009 & 0.0380 $\pm$ 0.0003 & \textbf{0.0373 $\pm$ 0.0003} & 0.0468 $\pm$ 0.0002 \\
KL        & 5.1057 $\pm$ 0.0867 & 5.6839 $\pm$ 0.3074 & 3.1319 $\pm$ 0.1293 & \textbf{2.9654 $\pm$ 0.1203} & 5.0349 $\pm$ 0.3012 \\
TV        & 0.6680 $\pm$ 0.0021 & 0.6309 $\pm$ 0.0068 & 0.4775 $\pm$ 0.0016 & \textbf{0.4726 $\pm$ 0.0018} & 0.4990 $\pm$ 0.0002 \\
classical fidelity  & 0.1987 $\pm$ 0.0026 & 0.2232 $\pm$ 0.0057 & 0.4578 $\pm$ 0.0028 & 0.4667 $\pm$ 0.0032 & \textbf{0.5065 $\pm$ 0.0024} \\
\midrule
\multicolumn{6}{l}{\textit{Validity metrics from \cite{gili_metrics_2024}}} \\
coverage     & \textbf{0.5620 $\pm$ 0.0235} & 0.4880 $\pm$ 0.0543 & 0.4373 $\pm$ 0.0206 & 0.4297 $\pm$ 0.0133 & 0.3533 $\pm$ 0.0390 \\
fidelity     & 0.0633 $\pm$ 0.0021 & 0.0217 $\pm$ 0.0047 & 0.0998 $\pm$ 0.0037 & \textbf{0.1129 $\pm$ 0.0036} & 0.0309 $\pm$ 0.0058 \\
rate         & \textbf{0.0403 $\pm$ 0.0012} & 0.0123 $\pm$ 0.0027 & 0.0270 $\pm$ 0.0011 & 0.0308 $\pm$ 0.0011 & 0.0008 $\pm$ 0.0001 \\
exploration  & \textbf{0.6370 $\pm$ 0.0028} & 0.5489 $\pm$ 0.0153 & 0.2711 $\pm$ 0.0035 & 0.2727 $\pm$ 0.0028 & 0.0277 $\pm$ 0.0016 \\
\midrule
\multicolumn{6}{l}{\textit{BAS metrics from \cite{benedetti_qcbm_benchmark_2019}}} \\
precision & 0.3382 $\pm$ 0.0043 & 0.3254 $\pm$ 0.0086 & 0.5327 $\pm$ 0.0028 & 0.5304 $\pm$ 0.0038 & \textbf{0.6963 $\pm$ 0.0032} \\
recall    & \textbf{0.7096 $\pm$ 0.0185} & 0.6665 $\pm$ 0.0346 & 0.6383 $\pm$ 0.0132 & 0.6334 $\pm$ 0.0086 & 0.5843 $\pm$ 0.0251 \\
qBAS      & 0.4562 $\pm$ 0.0051 & 0.4313 $\pm$ 0.0106 & 0.5797 $\pm$ 0.0053 & 0.5769 $\pm$ 0.0045 & \textbf{0.6307 $\pm$ 0.0161} \\
\bottomrule
\end{tabular*}
\end{table*}

\subsubsection{Threshold Sensitivity}
The metric-based extension is heuristically set by a threshold value. It is possible to define other rules to set this threshold, which raises the question of the sensitivity of this threshold. We carry out additional simulations with the same setup as in Sec.~\ref{sec:bas_training}, but varying threshold values. These values are chosen with a higher resolution where the number of additional connections is most sensitive to the threshold (see Fig.~\ref{fig:bas_threshold_curve}).

\begin{figure}[ht]
\centering
\includegraphics[width=1.0\columnwidth]{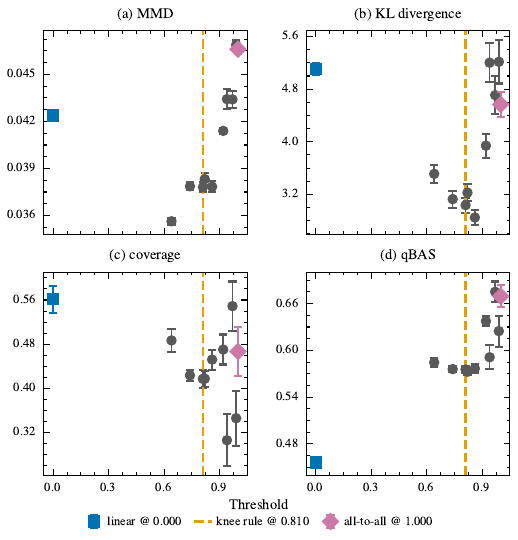}
\caption{Benchmark metrics vs. extension threshold on the BAS dataset: (a) MMD,
(b) KL, (c) coverage, (d) qBAS. Each point is the bootstrapped mean $\pm$ standard
error across 16 runs per threshold; threshold $=0$ coincides with the unextended
linear circuit and threshold $=1$ with all-to-all connectivity, the dashed line marks the threshold selected by the knee rule.}
\label{fig:threshold_sweep}
\end{figure}

The selected evaluation metrics for the threshold sweep in Fig.~\ref{fig:threshold_sweep} show that adding more connections does not simply make the model better or worse. There is a middle range of thresholds where it performs best. In this middle range, MMD and KL are lowest, meaning the model fits the full data distribution most closely, while both very low and very high thresholds fit it worse. Coverage behaves the opposite way: it is highest for the unextended circuit, drops in that same middle range, and only partly recovers as the threshold increases further. This reflects a tradeoff between fitting the target distribution well and still producing diverse, previously unseen valid patterns. The qBAS score stays roughly constant across the low and middle thresholds and increases further once the threshold approaches full connectivity. Overall, the automatically selected threshold lands in the region that fits the distribution well and avoids potential drops in coverage seen at higher connectivity, even though it does not give the highest possible qBAS score.

\subsection{Interest Rates}\label{sec:jgb_experiment}
\begin{figure}[ht]
\centering
\includegraphics[width=1.0\columnwidth]{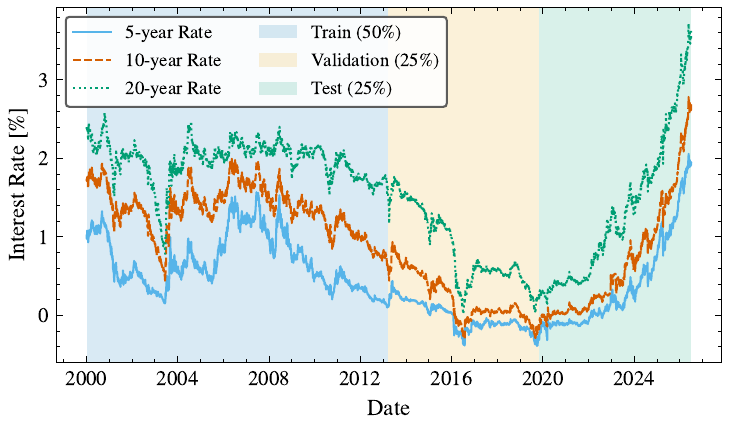}
\caption{Daily Japanese Government Bond (JGB) yield rates for the 5-, 10-, and 20-year maturities. Shading marks the chronological train/validation/test split, with training on the earliest period and testing on the most recent.
\label{fig:sr_raw_data}}
\end{figure}

We apply the metric-based extension heuristic to a financial dataset consisting of daily quotes of interest rates from Japanese government bonds~(JGB) to examine its practicality in real-world applications. Although the recent literature in generative modeling for finance mainly focused on experiments using a foreign exchange rates dataset~\cite{coyle_finance_2021,kondratyev_non_differentiable_2019,kondratyev_market_2019}, we consider the interest rates dataset a more suitable choice, because in contrast to foreign exchange rates, interest rates exhibit stronger correlations driven by the inherent structure of the interest rate curve~\cite{sadr_swaps_2009}. We consider three types of JGBs, which differ in their maturity periods of 5, 10, and 20 years. The dataset can be freely downloaded from the official website of the Ministry of Finance Japan and consists of daily observations of the interest rates of each JGB. We use a period from 2000-01-04 to 2026-06-30, as visualized in Fig.~\ref{fig:sr_raw_data}.

We preprocess the dataset as follows. First, we difference the interest rates time series $S_t$ by $D_t = S_t - S_{t-1}$, to represent the day-by-day changes and ensure stationarity. Second, since the QCBM represents data in binary format, as each qubit collapses to either 0 or 1 upon measurement, it is necessary to quantize the data distributions to a desired bit resolution. As commonly used in recent literature, we adopt the uniform ``minmax'' quantization~\cite{kondratyev_market_2019}. A decimal value $(x)_{10}$ is encoded into its $n$-bit binary equivalent $(x)_{2}$ by
\begin{equation}\label{eqn:dec_to_bin}
(x)_{2} = \left\lfloor (2^n - 1) \frac{(x)_{10} - x_{\text{min}}}{x_{\text{max}} - x_{\text{min}}} \right\rfloor,
\end{equation}
where $x_{\text{min}}$ and $x_{\text{max}}$ are the minimum and maximum values of the training split, so that no information from the validation or test split enters the encoding. The floor operation $\lfloor \cdot \rfloor$ ensures that the result remains an integer. Values outside $[x_{\text{min}}, x_{\text{max}}]$, which can only occur in the validation and test splits, are clipped onto the two extreme levels $0$ and $2^n-1$.

To finally decode the binary value back to its decimal representation, we apply the midpoint reconstruction rule
\begin{equation}\label{eqn:bin_to_dec}
(x)_{10} = x_{\text{min}} + \left((x)_{2} + \tfrac{1}{2}\right) \frac{x_{\text{max}} - x_{\text{min}}}{2^n - 1}.
\end{equation}
In our experiments, we restrict the algorithm to utilize 12 qubits, so each of the 3 features has to be quantized to a 4-bit resolution.

\subsubsection{Circuit Design}
\begin{figure}[ht]
\centering{\includegraphics[width=1.0\columnwidth]{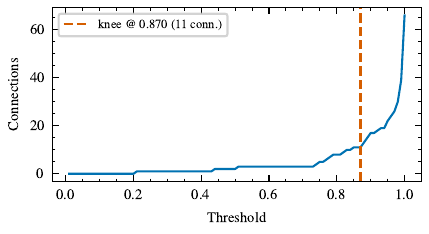}}
\caption{Number of added connections vs.\ threshold for the metric-based extension on the JGB dataset, using the variation-of-information distance between features. The dashed line marks the threshold used, adding $11$ connections, $3$ of which are already included in the linear baseline.
\label{fig:jgb_threshold}}
\end{figure}

We apply the metric-based extension heuristic via variation-of-information on the differenced and quantized interest rates distribution of the training set. The particular threshold was set by the knee rule, as shown in Fig.~\ref{fig:jgb_threshold}. As a good guess for the initial connections, we suggest establishing a linear connection between the bits that belong to one time-series feature, e.g. \{0,1\}, \{1,2\}, and \{2,3\}. The extended circuits are included in Fig.~\ref{fig:jgb_extensions_mmd}. The metric-based extension heuristic extends the circuit by eight additional gates, whereas Chow-Liu adds six. From visual inspection, we see that the first two qubits of each feature tend to be connected with each other: \{0,1\} of the 5-year rate with \{4,5\} of the 10-year rate, and \{4,5\} with \{8,9\} of the 20-year rate. Unlike in the BAS experiment, where the metric-based and Chow-Liu methods nearly recovered an identical circuit, here Chow-Liu adds different connections, such as \{2,6\}, \{6,10\}, \{8,10\} and \{8,11\}.

\subsubsection{Training}

\begin{figure*}[ht]
\centering
\includegraphics[width=1.0\textwidth]{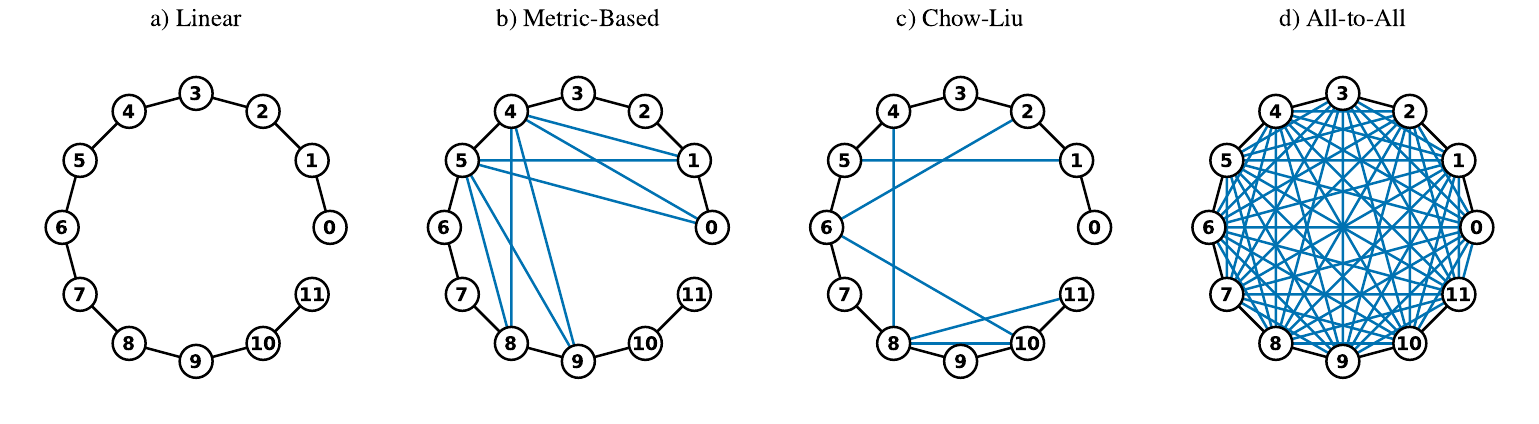}
\includegraphics[width=1.0\textwidth]{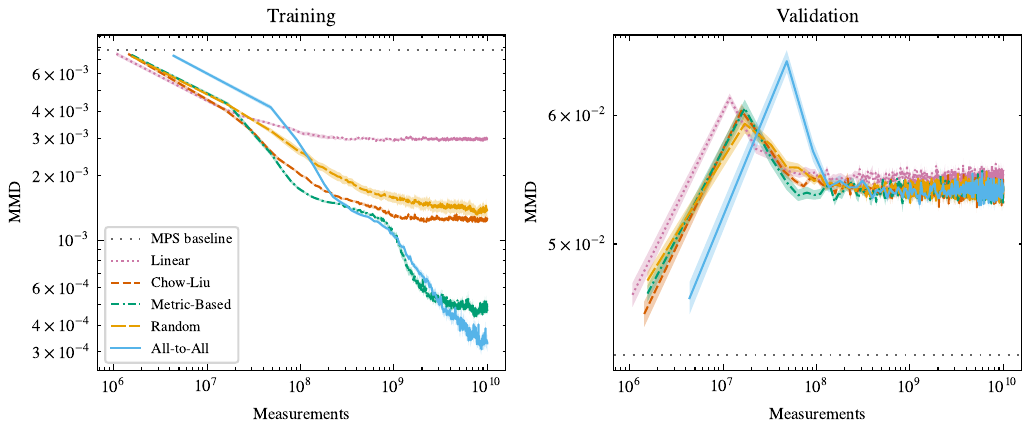}
\caption{Top: connectivity of the JGB circuit extensions. Black edges are the linear connectivity derived from the pretrained MPS model; blue edges are the added connections for (a) no extension, (b) metric-based, (c) Chow-Liu, and (d) all-to-all. Qubits are grouped by feature: $\{0,1,2,3\}$ encode the 5-year rate, $\{4,5,6,7\}$ the 10-year rate, and $\{8,9,10,11\}$ the 20-year rate. Bottom: MMD vs.\ cumulative measurements on the JGB training (left) and validation (right) splits, log-log scale, for the same five connectivities: linear, Chow-Liu, metric-based, random, and all-to-all. Lines show the bootstrapped mean and shaded bands show its standard deviation across 16 independent runs per connectivity; the grey dotted line is the MMD of the untrained, MPS-derived linear circuit.
\label{fig:jgb_extensions_mmd}}
\end{figure*}

The JGB simulations are carried out according to Appendix~\ref{app:hyperparameters}, sharing mostly the same parameters as the BAS simulations. We compare the MMD loss during training, evaluated after every tenth iteration on the training and the validation sets. The training and validation MMD curves in Fig.~\ref{fig:jgb_extensions_mmd} show that added connectivity improves the fit to the training data but does not transfer to unseen data. On the training split, all connectivities start from the untrained MPS baseline and then separate clearly, with all-to-all reaching the lowest loss, followed closely by the metric-based extension, while Chow-Liu and random settle at higher values and the linear circuit plateaus earliest and highest. Notably, the metric-based extension comes close to all-to-all while using far fewer parameters and connections, and it clearly outperforms the random extension despite both using the same number of added connections. The validation split shows that none of the trained models generalize well. Every connectivity decays into a common noisy plateau where the connectivities are no longer distinguishable. Throughout, the validation loss remains above the untrained MPS baseline, so training substantially reduces the training loss while leaving the validation loss worse than at initialization.

\subsubsection{Benchmarking}

Every run is evaluated at its final iteration with $10\,000$ measurements against the full dataset. Table~\ref{tab:jgb_benchmark} reports the mean and standard deviation of distribution distance metrics across all trained models for each connectivity via bootstrap resampling. With an equal number of parameters, the metric-based circuit compares favorably to the random one on all distributional metrics. The margin is clear for KL, TV, and classical fidelity, where it exceeds the bootstrap uncertainty, while the MMD values of the two are comparable within their error bars. The all-to-all circuit performs the best on all metrics except KL, where metric-based closely matches its performance. The Chow-Liu circuit performs similarly to the random connectivity, while the linear circuit is the worst overall.

\begin{table*}[ht]
\caption{Bootstrapped benchmark metrics (mean $\pm$ standard error across 16 runs per connectivity) on the JGB dataset. Bold marks the best value per metric row.}
\label{tab:jgb_benchmark}
\tabfont
\begin{tabular*}{\textwidth}{@{\extracolsep{\fill}}lccccc@{}}
\toprule
Metric & Linear & Chow-Liu & Metric-Based & Random & All-to-All \\
\midrule
\multicolumn{6}{l}{\textit{Setup}} \\
parameters      & 267 & 357 & 387 & 387 & 1{,}092 \\
extension gates  & 0 & 6 & 8 & 8 & 55 \\
measurements & 9{,}999{,}150{,}000 & 9{,}999{,}990{,}000 & 9{,}999{,}050{,}000 & 9{,}999{,}050{,}000 & 9{,}998{,}560{,}000 \\
iterations   & 9{,}345 & 6{,}993 & 6{,}451 & 6{,}451 & 2{,}288 \\
\midrule
\multicolumn{6}{l}{\textit{Distribution distance metrics}} \\
MMD       & 0.0092 $\pm$ 0.0004 & 0.0083 $\pm$ 0.0004 & 0.0075 $\pm$ 0.0003 & 0.0079 $\pm$ 0.0003 & \textbf{0.0071 $\pm$ 0.0004} \\
KL        & 1.4113 $\pm$ 0.0161 & 1.2341 $\pm$ 0.0171 & \textbf{0.5158 $\pm$ 0.0114} & 1.2108 $\pm$ 0.0446 & 0.5372 $\pm$ 0.0276 \\
TV        & 0.3498 $\pm$ 0.0037 & 0.2669 $\pm$ 0.0031 & 0.2206 $\pm$ 0.0031 & 0.2782 $\pm$ 0.0049 & \textbf{0.2053 $\pm$ 0.0033} \\
classical fidelity  & 0.6476 $\pm$ 0.0036 & 0.7138 $\pm$ 0.0018 & 0.8233 $\pm$ 0.0027 & 0.7006 $\pm$ 0.0070 & \textbf{0.8299 $\pm$ 0.0042} \\
\bottomrule
\end{tabular*}
\end{table*}

\begin{figure*}[ht]
    \centering{\includegraphics[width=1.0\textwidth]{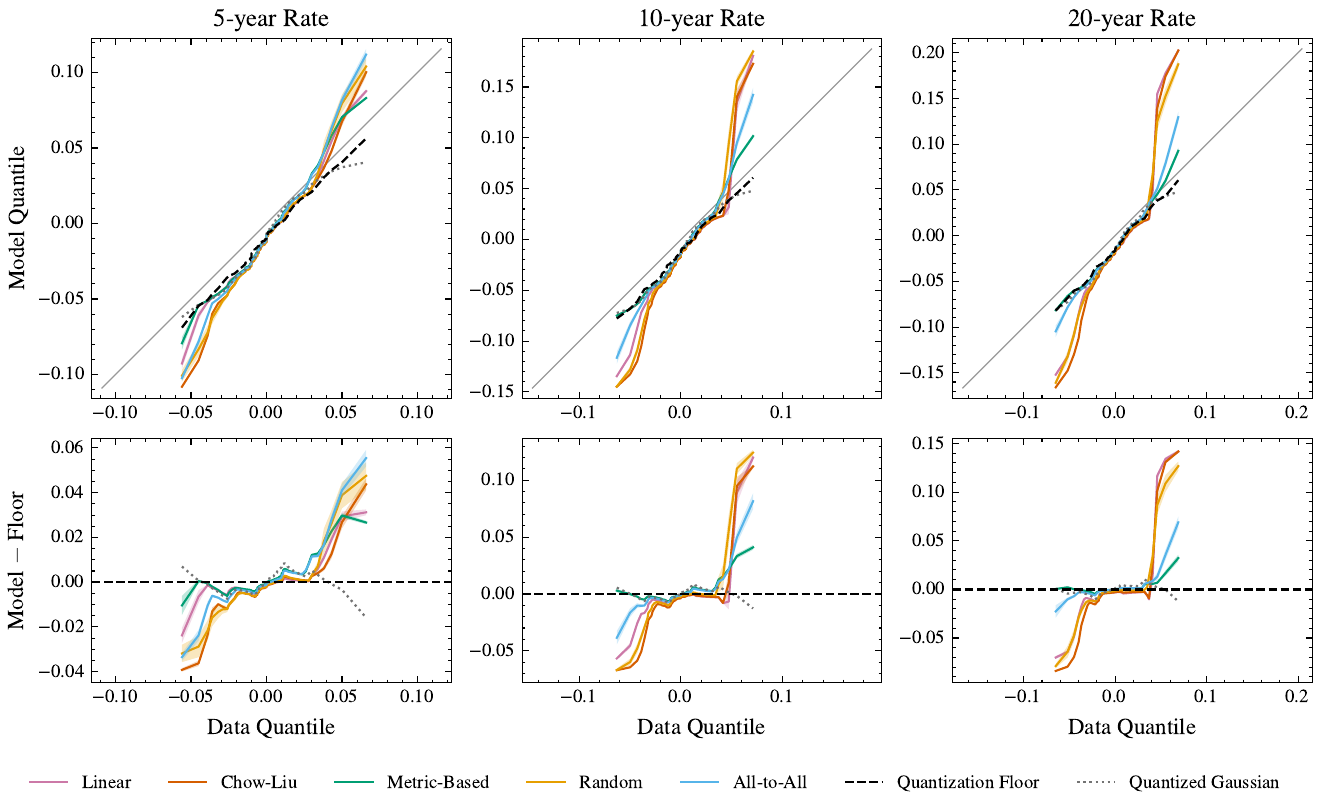}}
    \caption{Quantile-quantile plots of the generated JGB rate changes against the empirical data, one panel per maturity. Each band shows the bootstrapped mean and its standard error over all runs of one connectivity. The black dashed line is the quantization floor, and the grey dotted line a Gaussian fitted to the quantized data. The bottom row shows the same bands relative to the quantization floor.
    \label{fig:jgb_qq_bootstrap}}
\end{figure*}

Fig.~\ref{fig:jgb_qq_bootstrap} compares the marginal distributions, after decoding the generated binary samples into decimal values for each feature. Through binary decoding, a quantization error is introduced. We show this error as the quantization floor by plotting the encoded--decoded dataset against the original dataset. This floor is visibly separated from the diagonal line in the upper panel, which represents the fit to the distribution of the original dataset. The floor can only be brought substantially closer to the original dataset by increasing the qubit resolution per feature. The quantization floor indicates the best fit that our trained models are able to achieve, and is therefore shown separately in the lower panel.

Across the center of each marginal, all connectivities match the quantization floor reasonably close. The lines diverge increasingly in the distribution's tails, where every connectivity overshoots the empirical extremes to differing degrees. Form visual inspection, the metric-based circuit deviates least and remains closest to the floor across all maturities, whereas linear, random, and Chow-Liu are consistently further from the floor. All-to-all ranks second on the 10-year and 20-year rates, but shows a worse fit on the 5-year rate. Additionally, we compare a quantized Gaussian, which shows a closer fit to the floor than all our models, but notably diverges in the opposite direction at the tails. Overall, these results demonstrate that the all-to-all connected circuit fits the joint distribution better, while our proposed metric-based extension represents the marginals better.

%% file: content/4_discussion.tex
\section{Discussion}\label{sec:discussion}

In this work, we presented experimental results from training a QCBM with circuits that introduce varying degrees of inductive bias. We introduced a heuristic for designing a circuit under limited hardware resources and applied it to extend a circuit from a pretrained linear ansatz to a more connected problem-informed ansatz. Carrying out numerical simulations on both a toy problem (the BAS dataset) and a real-world financial dataset composed of JGB interest rates, we explored the effect of different circuit extensions on the subsequent retraining of the extended circuit.

On both datasets, the metric-based extension achieves a better distributional fit than random extensions using the same number of parameters, and therefore the same measurement cost per iteration. This difference suggests that the placement of the entangling gates, rather than the parameter count itself, drives the improvement in distributional fit. The comparison to Chow-Liu clarifies how it differs from our conceptually similar proposal. On BAS, both circuits nearly coincide, whereas on JGB they select different edges. Since the threshold rule is not restricted to a tree, it can admit loops, which appears to provide a stronger advantage on the real-world JGB dataset than on BAS. We find that denser connectivity does not translate directly into better models. The all-to-all circuit attains the lowest training loss in both experiments, but shows stronger signs of overfitting. On BAS, all-to-all therefore leads on precision-like metrics, while recording the worst coverage, exploration, and recall. The sparse problem-informed extensions achieve most of the distributional fit at a fraction of the parameter cost, without a collapse in diversity. On JGB, the same tension reappears in a different way. The all-to-all connected circuit fits the joint distribution better, while our metric-based extension shows a closer fit to the quantization floor across the marginal distributions. No single connectivity dominates every metric, which cautions against benchmarking circuit designs on a single figure of merit. 

Several limitations of our experiments warrant caution when interpreting these results. The most significant limitation concerns generalization: on both datasets, the validation loss stagnates near the untrained MPS baseline while the training loss continues to improve. Our results therefore establish a relative ordering between circuit designs, but do not demonstrate that the trained models generalize well in absolute terms. The financial experiment carries additional caveats, as the chronological split might place validation and test data in a different interest rate regime and thereby potentially violate the i.i.d.\ assumption. Further, temporal dependence and conditional structure remain outside the scope of a static joint-distribution model, like the QCBM we employed. The coarse per-feature quantization imposes a hard floor that dominates the achievable fit of the trained model. Moreover, our noiseless statevector simulation excludes both device noise and the routing overhead of the denser connectivities, which would penalize all-to-all more than the sparse extensions. 

Since the proposed extension uses parameterized SU(4) gates, trainability and barren plateau effects must be considered carefully. In this work, we mitigated this issue through three design choices. First, the circuit is initialized from a pretrained MPS rather than with random parameters. Second, the metric-based extension heuristic adds only a limited number of two-qubit gates selected according to feature similarity, thereby avoiding unnecessarily deep or highly expressive all-to-all circuits. Third, the added SU(4) gates are initialized close to the identity by sampling their parameters from a narrow normal distribution. These are intended to prevent the immediate generation of random global entanglement and to keep the circuit in a trainable regime. Despite the successful training in the investigated regimes, a systematic study of gradient variance and measurement complexity as a function of qubit number and circuit depth is left for future work.

Although the proposed metric-based extension performs better than random extensions in our setting, further experiments are required to determine whether this behavior persists for larger feature spaces and deeper circuits. In the worst case, the number of possible two-qubit connections grows quadratically with the number of qubits. The thresholding step in the proposed heuristic is therefore essential for selecting a sparse subset of feature-relevant connections and avoiding the all-to-all limit. For larger systems, additional constraints such as locality restrictions or hardware-aware edge pruning may be required, which we will address in future work.

On the theoretical side, we did not establish a formal connection between classical feature similarity, entanglement capacity, and the circuit's inductive bias. This is a fundamental gap that should be closed before further exploration of problem-informed circuits based on feature similarity. Moreover, better quantum circuit design is only valuable if the resulting models are competitive with classical alternatives -- a question our connectivity-only comparisons cannot can answer. Consequently, we consider systematic benchmarking against the strongest classical baselines the most important direction for the broader field of QML.

While challenges remain, the findings of this work provide an experimental basis for future research in circuit design for quantum generative modeling and its applications in finance and beyond.

%% file: content/acks.tex
M.M. acknowledges support from the PROMOS scholarship issued by DAAD and funded by the German Ministry of Education and Research, as well as the MIRAI scholarship issued and funded by RWTH Aachen University. J.K. acknowledges support by SIP Grant Number JPJ012367. This work was supported by MEXT Quantum Leap Flagship Program Grants No. JPMXS0118067285 and No. JPMXS0120319794. 

%% file: content/appendix.tex
\section{Simulation Hyperparameters}
\label{app:hyperparameters}

Table~\ref{tab:hyperparameters} lists the hyperparameters used for all experiments. Values are shared across both datasets unless stated otherwise.

\begin{table}[ht]
\centering
\footnotesize
\begin{tabular}{@{}ll@{}}
\toprule
Parameter & Value \\
\midrule
\multicolumn{2}{@{}l}{\textit{Dataset}} \\
BAS grid size / qubits          & $3\times3$ / 9 \\
JGB features / qubits           & 3 (5-, 10-, 20-year rates) / 12 \\
JGB quantizer                   & minmax (uniform bins) \\
Train / validation / test split & 0.5 / 0.25 / 0.25 \\
\addlinespace
\multicolumn{2}{@{}l}{\textit{MPS pretraining}} \\
Cutoff precision       & $5\times10^{-5}$ \\
Descent step length    & 0.05 \\
Descent steps          & 10 \\
Training loops         & 10 \\
\addlinespace
\multicolumn{2}{@{}l}{\textit{Circuit extension}} \\
Distance metric        & variation of information \\
Threshold rule         & knee \\
Gate per added edge    & SU(4), 15 parameters \\
\addlinespace
\multicolumn{2}{@{}l}{\textit{QCBM training}} \\
Loss function                             & MMD \\
MMD kernel bandwidths $\sigma$            & $\{0.25, 1.0, 4.0, 16.0\}$ \\
Gradient estimator                        & parameter shift, $\pm\pi/2$ \\
Optimizer                                 & Adam \\
Adam initial learning rate                & 0.01 \\
Adam $\beta_1$, $\beta_2$, $\epsilon$     & 0.9, 0.999, $10^{-8}$ \\
$N_{\text{shots}}$ per circuit evaluation & 2000 \\
Measurements per iteration                & $(2P+1)\times N_{\text{shots}}$, $P$ parameters \\
Measurement budget per run                & $1\times10^{10}$ \\
\addlinespace
\multicolumn{2}{@{}l}{\textit{Simulation}} \\
Simulator                  & Qiskit Aer, statevector, CPU \\
Runs per circuit extension & 16 \\
\addlinespace
\multicolumn{2}{@{}l}{\textit{Evaluation}} \\
Shots per benchmark sample & $10\,000$ \\
Bootstrap resamples        & 1000 \\
\bottomrule
\end{tabular}
\caption{Hyperparameters used for the MPS pretraining, circuit extension, QCBM training, and evaluation.}
\label{tab:hyperparameters}
\end{table}

%% file: ref_v2.bib
@article{kondratyev_market_2019,
  author = {Oleksiy Kondratyev and Christian Schwarz},
  title = {The Market Generator},
  journal = {SSRN},
  year = {2019},
  doi = {https://dx.doi.org/10.2139/ssrn.3384948}
}

@article{kondratyev_non_differentiable_2019,
  author = {Kondratyev, Alex},
  title = {Non-Differentiable Learning of Quantum Circuit {Born} Machine with Genetic Algorithm},
  journal = {Wilmott},
  volume = {2021},
  number = {114},
  pages = {50-61},
  doi = {https://doi.org/10.1002/wilm.10943},
  url = {https://onlinelibrary.wiley.com/doi/abs/10.1002/wilm.10943},
  year = {2021}
}

@article{coyle_born_2020,
  author = {Brian Coyle and Daniel Mills and Vincent Danos and Elham Kashefi},
  title = {The {Born} Supremacy: Quantum Advantage and Training of an {Ising} {Born} Machine},
  journal = {npj Quantum Inf},
  year = {2020},
  volume  =	6,
  number  = 60,
  doi     = {https://doi.org/10.1038/s41534-020-00288-9},
}

@article{liu_differentiable_2018,
  title = {Differentiable learning of quantum circuit {Born} machines},
  author = {Liu, Jin-Guo and Wang, Lei},
  journal = {Phys. Rev. A},
  volume = {98},
  issue = {6},
  pages = {062324},
  numpages = {9},
  year = {2018},
  month = {Dec},
  publisher = {American Physical Society},
  doi = {10.1103/PhysRevA.98.062324},
  url = {https://link.aps.org/doi/10.1103/PhysRevA.98.062324}
}

@proceedings{diederik_adam_2014,
  editor = {Diederik P. Kingma and Jimmy Lei Ba},
  title = {Adam: A Method for Stochastic Optimization},
  year = {2014},
  volume = 3,
  organization = {International Conference for Learning Representations},
  venue = {San Diego},
  eventdate = {2015},
  url = {https://doi.org/10.48550/arXiv.1412.6980}
}

@article{han_unsupervised_2018,
  title = {Unsupervised Generative Modeling Using Matrix Product States},
  author = {Han, Zhao-Yu and Wang, Jun and Fan, Heng and Wang, Lei and Zhang, Pan},
  journal = {Phys. Rev. X},
  volume = {8},
  issue = {3},
  pages = {031012},
  numpages = {13},
  year = {2018},
  month = {Jul},
  publisher = {American Physical Society},
  doi = {10.1103/PhysRevX.8.031012},
  url = {https://link.aps.org/doi/10.1103/PhysRevX.8.031012}
}

@online{han_mps_github_2018,
  title = {Unsupervised Generative Modeling Using Matrix Product States},
  author = {Han, Zhao-Yu and Wang, Jun and Fan, Heng and Wang, Lei and Zhang, Pan},
  url = {https://github.com/congzlwag/UnsupGenModbyMPS},
  urldate = {2024-12-20},
  year = {2018},
  organization = {GitHub, Commit 2fe52f2}
}

@article{rudolph_decomposition_2023,
  doi = {10.1088/2058-9565/ad04e6},
  url = {https://dx.doi.org/10.1088/2058-9565/ad04e6},
  year = {2023},
  month = {nov},
  publisher = {IOP Publishing},
  volume = {9},
  number = {1},
  pages = {015012},
  author = {Manuel S Rudolph and Jing Chen and Jacob Miller and Atithi Acharya and Alejandro Perdomo-Ortiz},
  title = {Decomposition of matrix product states into shallow quantum circuits},
  journal = {Quantum Science and Technology}
}

@article{ran_analytic_decomp_2020,
  title = {Encoding of matrix product states into quantum circuits of one- and two-qubit gates},
  author = {Ran, Shi-Ju},
  journal = {Phys. Rev. A},
  volume = {101},
  issue = {3},
  pages = {032310},
  numpages = {7},
  year = {2020},
  month = {Mar},
  publisher = {American Physical Society},
  doi = {10.1103/PhysRevA.101.032310},
  url = {https://link.aps.org/doi/10.1103/PhysRevA.101.032310}
}

@article{rudolph_synergistic_2023,
  author = {Manuel S. Rudolph and Jacob Miller and Danial Motlagh and Jing Chen and Atithi Acharya and Alejandro Perdomo-Ortiz},
  title = {Synergistic pretraining of parametrized quantum circuits via tensor networks.},
  journal = {Nat Commun},
  volume = {14},
  number = {8367},
  year = {2023},
  doi = {https://doi.org/10.1038/s41467-023-43908-6},
}

@article{rudolph_trainability_2024,
  author = {Manuel S. Rudolph and Sacha Lerch and Supanut Thanasilp and Oriel Kiss and Sofia Vallecorsa and Michele Grossi and Zoë Holmes},
  title = {Trainability barriers and opportunities in quantum generative modeling},
  journal = {npj Quantum Inf},
  volume = {10},
  number = {116},
  year = {2024},
  doi = {https://doi.org/10.1038/s41534-024-00902-0},
}

@article{benedetti_qcbm_benchmark_2019,
  author = {Marcello Benedetti and Delfina Garcia-Pintos and Oscar Perdomo and Vicente Leyton-Ortega and Yunseong Nam and Alejandro Perdomo-Ortiz},
  title = {A generative modeling approach benchmarking and training shallow quantum circuits},
  journal = {npj Quantum Inf},
  volume = {5},
  number = {45},
  year = {2019},
  doi = {https://doi.org/10.1038/s41534-019-0157-8},
}

@article{mcclean_barren_2018,
  author = {Jarrod R. McClean and Sergio Boixo and Vadim N. Smelyanskiy and Ryan Babbush and Hartmut Neven},
  title = {Barren plateaus in quantum neural network training landscapes},
  journal = { Nat Commun },
  volume = {9},
  number = {4812},
  year = {2018},
  doi = {https://doi.org/10.1038/s41467-018-07090-4},
}

@article{schuld_evaluating_2019,
  title = {Evaluating analytic gradients on quantum hardware},
  author = {Schuld, Maria and Bergholm, Ville and Gogolin, Christian and Izaac, Josh and Killoran, Nathan},
  journal = {Phys. Rev. A},
  volume = {99},
  issue = {3},
  pages = {032331},
  numpages = {7},
  year = {2019},
  month = {Mar},
  publisher = {American Physical Society},
  doi = {10.1103/PhysRevA.99.032331},
  url = {https://link.aps.org/doi/10.1103/PhysRevA.99.032331}
}

@article{du_expressive_2020,
  title = {Expressive power of parametrized quantum circuits},
  author = {Du, Yuxuan and Hsieh, Min-Hsiu and Liu, Tongliang and Tao, Dacheng},
  journal = {Phys. Rev. Res.},
  volume = {2},
  issue = {3},
  pages = {033125},
  numpages = {16},
  year = {2020},
  month = {Jul},
  publisher = {American Physical Society},
  doi = {10.1103/PhysRevResearch.2.033125},
  url = {https://link.aps.org/doi/10.1103/PhysRevResearch.2.033125}
}

@article{anderson_mmd_1994,
  title = {Two-Sample Test Statistics for Measuring Discrepancies Between Two Multivariate Probability Density Functions Using Kernel-Based Density Estimates},
  journal = {Journal of Multivariate Analysis},
  volume = {50},
  number = {1},
  pages = {41-54},
  year = {1994},
  issn = {0047-259X},
  doi = {https://doi.org/10.1006/jmva.1994.1033},
  url = {https://www.sciencedirect.com/science/article/pii/S0047259X84710335},
  author = {N.H. Anderson and P. Hall and D.M. Titterington},
}

@article{hofmann_kernel_2008,
 ISSN = {00905364},
 URL = {http://www.jstor.org/stable/25464664},
 author = {Thomas Hofmann and Bernhard Schölkopf and Alexander J. Smola},
 journal = {The Annals of Statistics},
 number = {3},
 pages = {1171--1220},
 publisher = {Institute of Mathematical Statistics},
 title = {Kernel Methods in Machine Learning},
 urldate = {2025-01-26},
 volume = {36},
 year = {2008}
}

@article{tucci_kak_2005,
  author = {Robert R. Tucci},
  title = {An Introduction to Cartan's KAK Decomposition for QC Programmers},
  journal = {arXiv preprint},
  year = {2005},
  url = {https://arxiv.org/abs/quant-ph/0507171}
}

@article{arabie_varinfo_1973,
  title = {Multidimensional scaling of measures of distance between partitions},
  journal = {Journal of Mathematical Psychology},
  volume = {10},
  number = {2},
  pages = {148-203},
  year = {1973},
  issn = {0022-2496},
  doi = {https://doi.org/10.1016/0022-2496(73)90012-6},
  url = {https://www.sciencedirect.com/science/article/pii/0022249673900126},
  author = {Phipps Arabie and Scott A. Boorman}
}

@article{hamilton_benchmark_2019,
  title = {Generative model benchmarks for superconducting qubits},
  author = {Hamilton, Kathleen E. and Dumitrescu, Eugene F. and Pooser, Raphael C.},
  journal = {Phys. Rev. A},
  volume = {99},
  issue = {6},
  pages = {062323},
  numpages = {9},
  year = {2019},
  month = {Jun},
  publisher = {American Physical Society},
  doi = {10.1103/PhysRevA.99.062323},
  url = {https://link.aps.org/doi/10.1103/PhysRevA.99.062323}
}

@book{sadr_swaps_2009,
  author = {Amir Sadr},
  title = {Interest Rate Swaps and Their Derivatives: A Practitioner's Guide},
  publisher = {John Wiley \& Sons, Ltd},
  address = {Chichester, UK},
  year = {2009},
  doi = {10.1002/9781118267967}
}

@article{harrow_quantum_2017,
	title = {Quantum computational supremacy},
	volume = {549},
	issn = {1476-4687},
	url = {https://doi.org/10.1038/nature23458},
	doi = {10.1038/nature23458},
	number = {7671},
	journal = {Nature},
	author = {Harrow, Aram W. and Montanaro, Ashley},
	month = sep,
	year = {2017},
	pages = {203--209},
}

@article{biamonte_quantum_2017,
	title = {Quantum machine learning},
	volume = {549},
	issn = {1476-4687},
	url = {https://doi.org/10.1038/nature23474},
	doi = {10.1038/nature23474},
	number = {7671},
	journal = {Nature},
	author = {Biamonte, Jacob and Wittek, Peter and Pancotti, Nicola and Rebentrost, Patrick and Wiebe, Nathan and Lloyd, Seth},
	month = sep,
	year = {2017},
	pages = {195--202},
}

@article{huang_quantum_2022,
  author = {Hsin-Yuan Huang  and Michael Broughton  and Jordan Cotler  and Sitan Chen  and Jerry Li  and Masoud Mohseni  and Hartmut Neven  and Ryan Babbush  and Richard Kueng  and John Preskill  and Jarrod R. McClean },
  title = {Quantum advantage in learning from experiments},
  journal = {Science},
  volume = {376},
  number = {6598},
  pages = {1182-1186},
  year = {2022},
  doi = {10.1126/science.abn7293},
  URL = {https://www.science.org/doi/abs/10.1126/science.abn7293},
  eprint = {https://www.science.org/doi/pdf/10.1126/science.abn7293},
}

@article{daley_practical_2022,
	title = {Practical quantum advantage in quantum simulation},
	volume = {607},
	issn = {1476-4687},
	url = {https://doi.org/10.1038/s41586-022-04940-6},
	doi = {10.1038/s41586-022-04940-6},
	number = {7920},
	journal = {Nature},
	author = {Daley, Andrew J. and Bloch, Immanuel and Kokail, Christian and Flannigan, Stuart and Pearson, Natalie and Troyer, Matthias and Zoller, Peter},
	month = jul,
	year = {2022},
	pages = {667--676},
}

@article{preskill_nisq_2018,
  doi = {10.22331/q-2018-08-06-79},
  url = {https://doi.org/10.22331/q-2018-08-06-79},
  title = {Quantum computing in the {NISQ} era and beyond},
  author = {Preskill, John},
  journal = {{Quantum}},
  issn = {2521-327X},
  publisher = {{Verein zur Förderung des Open Access Publizierens in den Quantenwissenschaften}},
  volume = {2},
  pages = {79},
  month = aug,
  year = {2018}
}

@article{coyle_finance_2021,
	title = {Quantum versus classical generative modelling in finance},
	volume = {6},
	url = {https://dx.doi.org/10.1088/2058-9565/abd3db},
	doi = {10.1088/2058-9565/abd3db},
	number = {2},
	journal = {Quantum Science and Technology},
	author = {Coyle, Brian and Henderson, Maxwell and Chan Jin Le, Justin and Kumar, Niraj and Paini, Marco and Kashefi, Elham},
	month = apr,
	year = {2021},
	note = {Publisher: IOP Publishing},
	pages = {024013},
}

@article{sweke_quantum_2021,
  doi = {10.22331/q-2021-03-23-417},
  url = {https://doi.org/10.22331/q-2021-03-23-417},
  title = {On the Quantum versus Classical Learnability of Discrete Distributions},
  author = {Sweke, Ryan and Seifert, Jean-Pierre and Hangleiter, Dominik and Eisert, Jens},
  journal = {Quantum},
  issn = {2521-327X},
  publisher = {Verein zur Förderung des Open Access Publizierens in den Quantenwissenschaften},
  volume = {5},
  pages = {417},
  month = mar,
  year = {2021}
}

@article{gao_quantumcorr_2022,
  title = {Enhancing Generative Models via Quantum Correlations},
  author = {Gao, Xun and Anschuetz, Eric R. and Wang, Sheng-Tao and Cirac, J. Ignacio and Lukin, Mikhail D.},
  journal = {Phys. Rev. X},
  volume = {12},
  issue = {2},
  pages = {021037},
  numpages = {26},
  year = {2022},
  month = {May},
  publisher = {American Physical Society},
  doi = {10.1103/PhysRevX.12.021037},
  url = {https://link.aps.org/doi/10.1103/PhysRevX.12.021037}
}

@article{vidal_efficient_2003,
  title = {Efficient Classical Simulation of Slightly Entangled Quantum Computations},
  author = {Vidal, Guifre},
  journal = {Phys. Rev. Lett.},
  volume = {91},
  issue = {14},
  pages = {147902},
  numpages = {4},
  year = {2003},
  month = {Oct},
  publisher = {American Physical Society},
  doi = {10.1103/PhysRevLett.91.147902},
  url = {https://link.aps.org/doi/10.1103/PhysRevLett.91.147902}
}

@article{schollwock_density_matrix_2005,
  title = {The density-matrix renormalization group},
  author = {Schollwöck, U.},
  journal = {Rev. Mod. Phys.},
  volume = {77},
  issue = {1},
  pages = {259--315},
  numpages = {0},
  year = {2005},
  month = {Apr},
  publisher = {American Physical Society},
  doi = {10.1103/RevModPhys.77.259},
  url = {https://link.aps.org/doi/10.1103/RevModPhys.77.259}
}

@article{chow_liu_tree_1968,
  author={Chow, C. and Liu, C.},
  journal={IEEE Transactions on Information Theory}, 
  title={Approximating discrete probability distributions with dependence trees}, 
  year={1968},
  volume={14},
  number={3},
  pages={462-467},
  doi={10.1109/TIT.1968.1054142}
}

@article{low_quantum_inference_2014,
  title = {Quantum inference on Bayesian networks},
  author = {Low, Guang Hao and Yoder, Theodore J. and Chuang, Isaac L.},
  journal = {Phys. Rev. A},
  volume = {89},
  issue = {6},
  pages = {062315},
  numpages = {7},
  year = {2014},
  month = {Jun},
  publisher = {American Physical Society},
  doi = {10.1103/PhysRevA.89.062315},
  url = {https://link.aps.org/doi/10.1103/PhysRevA.89.062315}
}

@article{sweke_stochastic_2020,
  doi = {10.22331/q-2020-08-31-314},
  url = {https://doi.org/10.22331/q-2020-08-31-314},
  title = {Stochastic gradient descent for hybrid quantum-classical optimization},
  author = {Sweke, Ryan and Wilde, Frederik and Meyer, Johannes and Schuld, Maria and Faehrmann, Paul K. and Meynard-Piganeau, Barth{\'{e}}l{\'{e}}my and Eisert, Jens},
  journal = {{Quantum}},
  issn = {2521-327X},
  publisher = {{Verein zur F{\"{o}}rderung des Open Access Publizierens in den Quantenwissenschaften}},
  volume = {4},
  pages = {314},
  month = aug,
  year = {2020}
}

@article{arrasmith_barren_2022,
  doi = {10.1088/2058-9565/ac7d06},
  url = {https://dx.doi.org/10.1088/2058-9565/ac7d06},
  year = {2022},
  month = {aug},
  publisher = {IOP Publishing},
  volume = {7},
  number = {4},
  pages = {045015},
  author = {Arrasmith, Andrew and Holmes, Zoë and Cerezo, M and Coles, Patrick J},
  title = {Equivalence of quantum barren plateaus to cost concentration and narrow gorges},
  journal = {Quantum Science and Technology},
}

@article{grant_initialization_2019,
  doi = {10.22331/q-2019-12-09-214},
  url = {https://doi.org/10.22331/q-2019-12-09-214},
  title = {An initialization strategy for addressing barren plateaus in parametrized quantum circuits},
  author = {Grant, Edward and Wossnig, Leonard and Ostaszewski, Mateusz and Benedetti, Marcello},
  journal = {{Quantum}},
  issn = {2521-327X},
  publisher = {Verein zur Foerderung des Open Access Publizierens in den Quantenwissenschaften},
  volume = {3},
  pages = {214},
  month = dec,
  year = {2019}
}

@article{patti_entanglement_2021,
  title = {Entanglement devised barren plateau mitigation},
  author = {Patti, Taylor L. and Najafi, Khadijeh and Gao, Xun and Yelin, Susanne F.},
  journal = {Phys. Rev. Res.},
  volume = {3},
  issue = {3},
  pages = {033090},
  numpages = {11},
  year = {2021},
  month = {Jul},
  publisher = {American Physical Society},
  doi = {10.1103/PhysRevResearch.3.033090},
  url = {https://link.aps.org/doi/10.1103/PhysRevResearch.3.033090}
}

@article{bako_problem_informed_2024,
  title={Problem-informed Graphical Quantum Generative Learning},
  author={Bak{\'o}, Bence and Nagy, D{\'a}niel TR and H{\'a}ga, P{\'e}ter and Kallus, Zs{\'o}fia and Zimbor{\'a}s, Zolt{\'a}n},
  journal={arXiv preprint arXiv:2405.14072},
  url={https://arxiv.org/abs/2405.14072}, 
  year={2024}
}

@book{koller_probabilistic_2009,
  author = {Daphne Koller and Nir Friedman},
  publisher = {The MIT Press},
  title = {Probabilistic Graphical Models: Principles and Techniques},
  year = {2009},
  address = {Cambridge, MA},
}

@article{cerezo_challenges_2022,
	title = {Challenges and opportunities in quantum machine learning},
	volume = {2},
	issn = {2662-8457},
	url = {https://doi.org/10.1038/s43588-022-00311-3},
	doi = {10.1038/s43588-022-00311-3},
	number = {9},
	journal = {Nature Computational Science},
	author = {Cerezo, M. and Verdon, Guillaume and Huang, Hsin-Yuan and Cincio, Lukasz and Coles, Patrick J.},
	month = sep,
	year = {2022},
	pages = {567--576},
}

@article{holmes_connecting_2022,
  title = {Connecting Ansatz Expressibility to Gradient Magnitudes and Barren Plateaus},
  author = {Holmes, Zo\"e and Sharma, Kunal and Cerezo, M. and Coles, Patrick J.},
  journal = {PRX Quantum},
  volume = {3},
  issue = {1},
  pages = {010313},
  numpages = {23},
  year = {2022},
  month = {Jan},
  publisher = {American Physical Society},
  doi = {10.1103/PRXQuantum.3.010313},
  url = {https://link.aps.org/doi/10.1103/PRXQuantum.3.010313}
}

@article{perdomo_ortiz_opportunities_2018,
doi = {10.1088/2058-9565/aab859},
url = {https://dx.doi.org/10.1088/2058-9565/aab859},
year = {2018},
month = {jun},
publisher = {IOP Publishing},
volume = {3},
number = {3},
pages = {030502},
author = {Perdomo-Ortiz, Alejandro and Benedetti, Marcello and Realpe-Gómez, John and Biswas, Rupak},
title = {Opportunities and challenges for quantum-assisted machine learning in near-term quantum computers},
journal = {Quantum Science and Technology},
}

@article{schuld_introduction_2015,
author = {Maria Schuld, Ilya Sinayskiy and Francesco Petruccione},
title = {An introduction to quantum machine learning},
journal = {Contemporary Physics},
volume = {56},
number = {2},
pages = {172--185},
year = {2015},
publisher = {Taylor \& Francis},
doi = {10.1080/00107514.2014.964942},
URL = {https://doi.org/10.1080/00107514.2014.964942},
eprint = {https://doi.org/10.1080/00107514.2014.964942},
}

@article{huang_power_2021,
	title = {Power of data in quantum machine learning},
	volume = {12},
	issn = {2041-1723},
	url = {https://doi.org/10.1038/s41467-021-22539-9},
	doi = {10.1038/s41467-021-22539-9},
	number = {1},
	journal = {Nature Communications},
	author = {Huang, Hsin-Yuan and Broughton, Michael and Mohseni, Masoud and Babbush, Ryan and Boixo, Sergio and Neven, Hartmut and McClean, Jarrod R.},
	month = may,
	year = {2021},
	pages = {2631},
}

@article{lloyd_quantum_2014,
	title = {Quantum principal component analysis},
	volume = {10},
	issn = {1745-2481},
	url = {https://doi.org/10.1038/nphys3029},
	doi = {10.1038/nphys3029},
	number = {9},
	journal = {Nature Physics},
	author = {Lloyd, Seth and Mohseni, Masoud and Rebentrost, Patrick},
	month = sep,
	year = {2014},
	pages = {631--633},
}

@article{harrow_linear_2009,
  title = {Quantum Algorithm for Linear Systems of Equations},
  author = {Harrow, Aram W. and Hassidim, Avinatan and Lloyd, Seth},
  journal = {Phys. Rev. Lett.},
  volume = {103},
  issue = {15},
  pages = {150502},
  numpages = {4},
  year = {2009},
  month = {Oct},
  publisher = {American Physical Society},
  doi = {10.1103/PhysRevLett.103.150502},
  url = {https://link.aps.org/doi/10.1103/PhysRevLett.103.150502}
}

@article{anschuetz_interpretable_2023,
  title = {Interpretable Quantum Advantage in Neural Sequence Learning},
  author = {Anschuetz, Eric R. and Hu, Hong-Ye and Huang, Jin-Long and Gao, Xun},
  journal = {PRX Quantum},
  volume = {4},
  issue = {2},
  pages = {020338},
  numpages = {22},
  year = {2023},
  month = {Jun},
  publisher = {American Physical Society},
  doi = {10.1103/PRXQuantum.4.020338},
  url = {https://link.aps.org/doi/10.1103/PRXQuantum.4.020338}
}

@article{gili_generalize_2023,
  title={Do quantum circuit {Born} machines generalize?},
  author={Gili, Kaitlin and Hibat-Allah, Mohamed and Mauri, Marta and Ballance, Chris and Perdomo-Ortiz, Alejandro},
  journal={Quantum Science and Technology},
  volume={8},
  number={3},
  pages={035021},
  year={2023},
  publisher={IOP Publishing},
  doi={10.1088/2058-9565/acd578}
}

@article{gili_metrics_2024,
  title        = {Generalization metrics for practical quantum advantage in generative models},
  author       = {Gili, Kaitlin and Mauri, Marta and Perdomo-Ortiz, Alejandro},
  journal      = {Phys. Rev. Applied},
  volume       = {21},
  pages        = {044032},
  year         = {2024},
  publisher    = {American Physical Society},
  doi          = {10.1103/PhysRevApplied.21.044032},
}

@misc{farhi_qaoa_2019,
      title={Quantum Supremacy through the Quantum Approximate Optimization Algorithm}, 
      author={Edward Farhi and Aram W Harrow},
      year={2019},
      eprint={1602.07674},
      archivePrefix={arXiv},
      primaryClass={quant-ph},
      url={https://arxiv.org/abs/1602.07674}, 
}

@article{bharti_nisq_2022,
  title = {Noisy intermediate-scale quantum algorithms},
  author = {Bharti, Kishor and Cervera-Lierta, Alba and Kyaw, Thi Ha and Haug, Tobias and Alperin-Lea, Sumner and Anand, Abhinav and Degroote, Matthias and Heimonen, Hermanni and Kottmann, Jakob S. and Menke, Tim and Mok, Wai-Keong and Sim, Sukin and Kwek, Leong-Chuan and Aspuru-Guzik, Al\'an},
  journal = {Rev. Mod. Phys.},
  volume = {94},
  issue = {1},
  pages = {015004},
  numpages = {69},
  year = {2022},
  month = {Feb},
  publisher = {American Physical Society},
  doi = {10.1103/RevModPhys.94.015004},
  url = {https://link.aps.org/doi/10.1103/RevModPhys.94.015004}
}

@article{aaronson_read_2015,
	title = {Read the fine print},
	volume = {11},
	issn = {1745-2481},
	url = {https://doi.org/10.1038/nphys3272},
	doi = {10.1038/nphys3272},
	number = {4},
	journal = {Nature Physics},
	author = {Aaronson, Scott},
	month = apr,
	year = {2015},
	pages = {291--293},
}

@book{national_academies_of_sciences_quantum_2019,
  title={Quantum computing: progress and prospects},
  author={Horowitz, Mark and Grumbling, Emily},
  year={2019},
  publisher={National Academies Press},
  address = {Washington, DC} 
}

@article{schuld_quest_2014,
	title = {The quest for a {Quantum} {Neural} {Network}},
	volume = {13},
	issn = {1573-1332},
	url = {https://doi.org/10.1007/s11128-014-0809-8},
	doi = {10.1007/s11128-014-0809-8},
	number = {11},
	journal = {Quantum Information Processing},
	author = {Schuld, Maria and Sinayskiy, Ilya and Petruccione, Francesco},
	month = nov,
	year = {2014},
	pages = {2567--2586},
}

@article{tian_recent_2022,
  title={Recent advances for quantum neural networks in generative learning},
  author={Tian, Jinkai and Sun, Xiaoyu and Du, Yuxuan and Zhao, Shanshan and Liu, Qing and Zhang, Kaining and Yi, Wei and Huang, Wanrong and Wang, Chaoyue and Wu, Xingyao and others},
  journal={IEEE Transactions on Pattern Analysis and Machine Intelligence},
  volume={45},
  number={10},
  pages={12321--12340},
  year={2023},
  publisher={IEEE}
}

@ARTICLE{bond_taylor_deep_2022,
  author={Bond-Taylor, Sam and Leach, Adam and Long, Yang and Willcocks, Chris G.},
  journal={IEEE Transactions on Pattern Analysis and Machine Intelligence}, 
  title={Deep Generative Modelling: A Comparative Review of {VAEs}, {GANs}, Normalizing Flows, Energy-Based and Autoregressive Models}, 
  year={2022},
  volume={44},
  number={11},
  pages={7327-7347},
  doi={10.1109/TPAMI.2021.3116668}}

@misc{terhal_adaptive_2004,
      title={Adaptive Quantum Computation, Constant Depth Quantum Circuits and Arthur-Merlin Games}, 
      author={Barbara M. Terhal and David P. DiVincenzo},
      year={2004},
      eprint={quant-ph/0205133},
      archivePrefix={arXiv},
      primaryClass={quant-ph},
      url={https://arxiv.org/abs/quant-ph/0205133}, 
}

@inproceedings{aaronson_linear_optics_2010,
  title={The computational complexity of linear optics},
  author={Aaronson, Scott and Arkhipov, Alex},
  booktitle={Proceedings of the forty-third annual ACM symposium on Theory of computing},
  pages={333--342},
  year={2011}
}

@article{arute_quantum_2019,
	title = {Quantum supremacy using a programmable superconducting processor},
	volume = {574},
	issn = {1476-4687},
	url = {https://doi.org/10.1038/s41586-019-1666-5},
	doi = {10.1038/s41586-019-1666-5},
	number = {7779},
	journal = {Nature},
	author = {Arute, Frank and Arya, Kunal and Babbush, Ryan and Bacon, Dave and Bardin, Joseph C. and Barends, Rami and Biswas, Rupak and Boixo, Sergio and Brandao, Fernando G. S. L. and Buell, David A. and Burkett, Brian and Chen, Yu and Chen, Zijun and Chiaro, Ben and Collins, Roberto and Courtney, William and Dunsworth, Andrew and Farhi, Edward and Foxen, Brooks and Fowler, Austin and Gidney, Craig and Giustina, Marissa and Graff, Rob and Guerin, Keith and Habegger, Steve and Harrigan, Matthew P. and Hartmann, Michael J. and Ho, Alan and Hoffmann, Markus and Huang, Trent and Humble, Travis S. and Isakov, Sergei V. and Jeffrey, Evan and Jiang, Zhang and Kafri, Dvir and Kechedzhi, Kostyantyn and Kelly, Julian and Klimov, Paul V. and Knysh, Sergey and Korotkov, Alexander and Kostritsa, Fedor and Landhuis, David and Lindmark, Mike and Lucero, Erik and Lyakh, Dmitry and Mandrà, Salvatore and McClean, Jarrod R. and McEwen, Matthew and Megrant, Anthony and Mi, Xiao and Michielsen, Kristel and Mohseni, Masoud and Mutus, Josh and Naaman, Ofer and Neeley, Matthew and Neill, Charles and Niu, Murphy Yuezhen and Ostby, Eric and Petukhov, Andre and Platt, John C. and Quintana, Chris and Rieffel, Eleanor G. and Roushan, Pedram and Rubin, Nicholas C. and Sank, Daniel and Satzinger, Kevin J. and Smelyanskiy, Vadim and Sung, Kevin J. and Trevithick, Matthew D. and Vainsencher, Amit and Villalonga, Benjamin and White, Theodore and Yao, Z. Jamie and Yeh, Ping and Zalcman, Adam and Neven, Hartmut and Martinis, John M.},
	month = oct,
	year = {2019},
	pages = {505--510},
}

@article{madsen_quantum_2022,
	title = {Quantum computational advantage with a programmable photonic processor},
	volume = {606},
	issn = {1476-4687},
	url = {https://doi.org/10.1038/s41586-022-04725-x},
	doi = {10.1038/s41586-022-04725-x},
	number = {7912},
	journal = {Nature},
	author = {Madsen, Lars S. and Laudenbach, Fabian and Askarani, Mohsen Falamarzi. and Rortais, Fabien and Vincent, Trevor and Bulmer, Jacob F. F. and Miatto, Filippo M. and Neuhaus, Leonhard and Helt, Lukas G. and Collins, Matthew J. and Lita, Adriana E. and Gerrits, Thomas and Nam, Sae Woo and Vaidya, Varun D. and Menotti, Matteo and Dhand, Ish and Vernon, Zachary and Quesada, Nicolás and Lavoie, Jonathan},
	month = jun,
	year = {2022},
	pages = {75--81},
}

@article{lloyd_qgan_2018,
  title = {Quantum Generative Adversarial Learning},
  author = {Lloyd, Seth and Weedbrook, Christian},
  journal = {Phys. Rev. Lett.},
  volume = {121},
  issue = {4},
  pages = {040502},
  numpages = {5},
  year = {2018},
  month = {Jul},
  publisher = {American Physical Society},
  doi = {10.1103/PhysRevLett.121.040502},
  url = {https://link.aps.org/doi/10.1103/PhysRevLett.121.040502}
}

@article{dallaire_demers_qgan_2018,
  title = {Quantum generative adversarial networks},
  author = {Dallaire-Demers, Pierre-Luc and Killoran, Nathan},
  journal = {Phys. Rev. A},
  volume = {98},
  issue = {1},
  pages = {012324},
  numpages = {8},
  year = {2018},
  month = {Jul},
  publisher = {American Physical Society},
  doi = {10.1103/PhysRevA.98.012324},
  url = {https://link.aps.org/doi/10.1103/PhysRevA.98.012324}
}

@article{amin_qbm_2018,
  title = {Quantum {Boltzmann} Machine},
  author = {Amin, Mohammad H. and Andriyash, Evgeny and Rolfe, Jason and Kulchytskyy, Bohdan and Melko, Roger},
  journal = {Phys. Rev. X},
  volume = {8},
  issue = {2},
  pages = {021050},
  numpages = {11},
  year = {2018},
  month = {May},
  publisher = {American Physical Society},
  doi = {10.1103/PhysRevX.8.021050},
  url = {https://link.aps.org/doi/10.1103/PhysRevX.8.021050}
}

@article{zoufal_variational_2021,
	title = {Variational quantum {Boltzmann} machines},
	volume = {3},
	issn = {2524-4914},
	url = {https://doi.org/10.1007/s42484-020-00033-7},
	doi = {10.1007/s42484-020-00033-7},
	number = {1},
	journal = {Quantum Machine Intelligence},
	author = {Zoufal, Christa and Lucchi, Aurélien and Woerner, Stefan},
	month = feb,
	year = {2021},
	pages = {7},
}

@article{larocca_diagnosing_2022,
  doi = {10.22331/q-2022-09-29-824},
  url = {https://doi.org/10.22331/q-2022-09-29-824},
  title = {Diagnosing Barren Plateaus with Tools from Quantum Optimal Control},
  author = {Larocca, Martin and Czarnik, Piotr and Sharma, Kunal and Muraleedharan, Gopikrishnan and Coles, Patrick J. and Cerezo, M.},
  journal = {{Quantum}},
  issn = {2521-327X},
  publisher = {{Verein zur F{\"{o}}rderung des Open Access Publizierens in den Quantenwissenschaften}},
  volume = {6},
  pages = {824},
  month = sep,
  year = {2022}
}

@article{cerezo_higher_2021,
	title = {Higher order derivatives of quantum neural networks with barren plateaus},
	volume = {6},
	url = {https://dx.doi.org/10.1088/2058-9565/abf51a},
	doi = {10.1088/2058-9565/abf51a},
	number = {3},
	journal = {Quantum Science and Technology},
	author = {Cerezo, M and Coles, Patrick J},
	month = jun,
	year = {2021},
	note = {Publisher: IOP Publishing},
	pages = {035006},
}

@article{arrasmith_effect_2021,
  doi = {10.22331/q-2021-10-05-558},
  url = {https://doi.org/10.22331/q-2021-10-05-558},
  title = {Effect of barren plateaus on gradient-free optimization},
  author = {Arrasmith, Andrew and Cerezo, M. and Czarnik, Piotr and Cincio, Lukasz and Coles, Patrick J.},
  journal = {{Quantum}},
  issn = {2521-327X},
  publisher = {{Verein zur F{\"{o}}rderung des Open Access Publizierens in den Quantenwissenschaften}},
  volume = {5},
  pages = {558},
  month = oct,
  year = {2021}
}

@article{holmes_scramblers_2021,
  title = {Barren Plateaus Preclude Learning Scramblers},
  author = {Holmes, Zo\"e and Arrasmith, Andrew and Yan, Bin and Coles, Patrick J. and Albrecht, Andreas and Sornborger, Andrew T.},
  journal = {Phys. Rev. Lett.},
  volume = {126},
  issue = {19},
  pages = {190501},
  numpages = {7},
  year = {2021},
  month = {May},
  publisher = {American Physical Society},
  doi = {10.1103/PhysRevLett.126.190501},
  url = {https://link.aps.org/doi/10.1103/PhysRevLett.126.190501}
}

@article{zhao_barren_2021,
  doi = {10.22331/q-2021-06-04-466},
  url = {https://doi.org/10.22331/q-2021-06-04-466},
  title = {Analyzing the barren plateau phenomenon in training quantum neural networks with the {ZX}-calculus},
  author = {Zhao, Chen and Gao, Xiao-Shan},
  journal = {{Quantum}},
  issn = {2521-327X},
  publisher = {{Verein zur F{\"{o}}rderung des Open Access Publizierens in den Quantenwissenschaften}},
  volume = {5},
  pages = {466},
  month = jun,
  year = {2021}
}

@article{thanasilp_exponential_2024,
  title={Exponential concentration in quantum kernel methods},
  author={Thanasilp, Supanut and Wang, Samson and Cerezo, Marco and Holmes, Zo{\"e}},
  journal={Nature communications},
  volume={15},
  number={1},
  pages={5200},
  year={2024},
  publisher={Nature Publishing Group UK London}
}

@article{ragone_lie_2024,
	title = {A {Lie} algebraic theory of barren plateaus for deep parameterized quantum circuits},
	volume = {15},
	issn = {2041-1723},
	url = {https://doi.org/10.1038/s41467-024-49909-3},
	doi = {10.1038/s41467-024-49909-3},
	number = {1},
	journal = {Nature Communications},
        author={Ragone, Michael and Bakalov, Bojko N and Sauvage, Fr{\'e}d{\'e}ric and Kemper, Alexander F and Ortiz Marrero, Carlos and Larocca, Mart{\'\i}n and Cerezo, M},
	month = aug,
	year = {2024},
	pages = {7172},
}

@article{fontana_characterizing_2024,
	title = {Characterizing barren plateaus in quantum ans\"{a}tze with the adjoint representation},
	volume = {15},
	issn = {2041-1723},
	url = {https://doi.org/10.1038/s41467-024-49910-w},
	doi = {10.1038/s41467-024-49910-w},
	number = {1},
	journal = {Nature Communications},
	author = {Fontana, Enrico and Herman, Dylan and Chakrabarti, Shouvanik and Kumar, Niraj and Yalovetzky, Romina and Heredge, Jamie and Sureshbabu, Shree Hari and Pistoia, Marco},
	month = aug,
	year = {2024},
	pages = {7171},
}

@article{anschuetz_quantum_2022,
	title = {Quantum variational algorithms are swamped with traps},
	volume = {13},
	issn = {2041-1723},
	url = {https://doi.org/10.1038/s41467-022-35364-5},
	doi = {10.1038/s41467-022-35364-5},
	number = {1},
	journal = {Nature Communications},
	author = {Anschuetz, Eric R. and Kiani, Bobak T.},
	month = dec,
	year = {2022},
	pages = {7760},
}

@misc{verdon_learning_2019,
      title={Learning to learn with quantum neural networks via classical neural networks}, 
      author={Guillaume Verdon and Michael Broughton and Jarrod R. McClean and Kevin J. Sung and Ryan Babbush and Zhang Jiang and Hartmut Neven and Masoud Mohseni},
      year={2019},
      eprint={1907.05415},
      archivePrefix={arXiv},
      primaryClass={quant-ph},
      url={https://arxiv.org/abs/1907.05415}, 
}

@article{larocca_group_2022,
  title = {Group-Invariant Quantum Machine Learning},
  author = {Larocca, Mart\'{\i}n and Sauvage, Fr\'ed\'eric and Sbahi, Faris M. and Verdon, Guillaume and Coles, Patrick J. and Cerezo, M.},
  journal = {PRX Quantum},
  volume = {3},
  issue = {3},
  pages = {030341},
  numpages = {25},
  year = {2022},
  month = {Sep},
  publisher = {American Physical Society},
  doi = {10.1103/PRXQuantum.3.030341},
  url = {https://link.aps.org/doi/10.1103/PRXQuantum.3.030341}
}

@article{sharma_trainability_2022,
  title = {Trainability of Dissipative Perceptron-Based Quantum Neural Networks},
  author = {Sharma, Kunal and Cerezo, M. and Cincio, Lukasz and Coles, Patrick J.},
  journal = {Phys. Rev. Lett.},
  volume = {128},
  issue = {18},
  pages = {180505},
  numpages = {7},
  year = {2022},
  month = {May},
  publisher = {American Physical Society},
  doi = {10.1103/PhysRevLett.128.180505},
  url = {https://link.aps.org/doi/10.1103/PhysRevLett.128.180505}
}

@article{cerezo_cost_2021,
	title = {Cost function dependent barren plateaus in shallow parametrized quantum circuits},
	volume = {12},
	issn = {2041-1723},
	url = {https://doi.org/10.1038/s41467-021-21728-w},
	doi = {10.1038/s41467-021-21728-w},
	number = {1},
	journal = {Nature Communications},
	author = {Cerezo, M. and Sone, Akira and Volkoff, Tyler and Cincio, Lukasz and Coles, Patrick J.},
	month = mar,
	year = {2021},
	pages = {1791},
}

@article{uvarov_barren_2021,
	title = {On barren plateaus and cost function locality in variational quantum algorithms},
	volume = {54},
	url = {https://dx.doi.org/10.1088/1751-8121/abfac7},
	doi = {10.1088/1751-8121/abfac7},
	number = {24},
	journal = {Journal of Physics A: Mathematical and Theoretical},
	author = {Uvarov, A V and Biamonte, J D},
	month = may,
	year = {2021},
	note = {Publisher: IOP Publishing},
	pages = {245301},
}

@article{marrero_entanglement_2021,
  title = {Entanglement-Induced Barren Plateaus},
  author = {Ortiz Marrero, Carlos and Kieferov\'a, M\'aria and Wiebe, Nathan},
  journal = {PRX Quantum},
  volume = {2},
  issue = {4},
  pages = {040316},
  numpages = {16},
  year = {2021},
  month = {Oct},
  publisher = {American Physical Society},
  doi = {10.1103/PRXQuantum.2.040316},
  url = {https://link.aps.org/doi/10.1103/PRXQuantum.2.040316}
}

@article{ho_no_free_lunch_2002,
	  title={Simple explanation of the no-free-lunch theorem and its implications},
  author={Ho, Yu-Chi and Pepyne, David L},
  journal={Journal of optimization theory and applications},
  volume={115},
  pages={549--570},
  year={2002},
  publisher={Springer}
}

@misc{poland_no_free_lunch_2020,
      title={No Free Lunch for Quantum Machine Learning}, 
      author={Kyle Poland and Kerstin Beer and Tobias J. Osborne},
      year={2020},
      eprint={2003.14103},
      archivePrefix={arXiv},
      primaryClass={quant-ph},
      url={https://arxiv.org/abs/2003.14103}, 
}

@article{sharma_no_free_lunch_2022,
  title = {Reformulation of the No-Free-Lunch Theorem for Entangled Datasets},
  author = {Sharma, Kunal and Cerezo, M. and Holmes, Zo\"e and Cincio, Lukasz and Sornborger, Andrew and Coles, Patrick J.},
  journal = {Phys. Rev. Lett.},
  volume = {128},
  issue = {7},
  pages = {070501},
  numpages = {7},
  year = {2022},
  month = {Feb},
  publisher = {American Physical Society},
  doi = {10.1103/PhysRevLett.128.070501},
  url = {https://link.aps.org/doi/10.1103/PhysRevLett.128.070501}
}

@article{meyer_exploiting_2023,
  title = {Exploiting Symmetry in Variational Quantum Machine Learning},
  author = {Meyer, Johannes Jakob and Mularski, Marian and Gil-Fuster, Elies and Mele, Antonio Anna and Arzani, Francesco and Wilms, Alissa and Eisert, Jens},
  journal = {PRX Quantum},
  volume = {4},
  issue = {1},
  pages = {010328},
  numpages = {27},
  year = {2023},
  month = {Mar},
  publisher = {American Physical Society},
  doi = {10.1103/PRXQuantum.4.010328},
  url = {https://link.aps.org/doi/10.1103/PRXQuantum.4.010328}
}

@article{zheng_speeding_2023,
  title = {Speeding Up Learning Quantum States Through Group Equivariant Convolutional Quantum Ans\"atze},
  author = {Zheng, Han and Li, Zimu and Liu, Junyu and Strelchuk, Sergii and Kondor, Risi},
  journal = {PRX Quantum},
  volume = {4},
  issue = {2},
  pages = {020327},
  numpages = {18},
  year = {2023},
  month = {May},
  publisher = {American Physical Society},
  doi = {10.1103/PRXQuantum.4.020327},
  url = {https://link.aps.org/doi/10.1103/PRXQuantum.4.020327}
}

@article{bowles_contextuality_2023,
  title={Contextuality and inductive bias in quantum machine learning},
  author={Bowles, Joseph and Wright, Victoria J and Farkas, M{\'a}t{\'e} and Killoran, Nathan and Schuld, Maria},
  journal={arXiv preprint arXiv:2302.01365},
  year={2023}
}

@article{borujeni_bayesian_2021,
  title={{Quantum circuit representation of Bayesian networks}},
  author={Borujeni, Sima E and Nannapaneni, Saideep and Nguyen, Nam H and Behrman, Elizabeth C and Steck, James E},
  journal={Expert Systems with Applications},
  volume={176},
  pages={114768},
  year={2021},
  publisher={Elsevier}
}

@article{paeckel2019time,
  title={Time-evolution methods for matrix-product states},
  author={Paeckel, Sebastian and K{\"o}hler, Thomas and Swoboda, Andreas and Manmana, Salvatore R and Schollw{\"o}ck, Ulrich and Hubig, Claudius},
  journal={Annals of Physics},
  volume={411},
  pages={167998},
  year={2019},
  publisher={Elsevier}
}
